\documentclass[journal,preprint]{vgtc}                
\vgtcinsertpkg
\onlineid{1407}
\category{Research}

\usepackage{balance}
\usepackage[utf8]{inputenc}

\usepackage{float}

\makeatletter
\newcommand{\printfnsymbol}[1]{%
  \textsuperscript{\@fnsymbol{#1}}%
}
\makeatother

\usepackage[utf8]{inputenc}
\usepackage[T1]{fontenc}
\usepackage{microtype}                 
\PassOptionsToPackage{warn}{textcomp}  
\usepackage{textcomp}                  
\usepackage{mathptmx}                  
\usepackage{times}                     
\usepackage{adjustbox}
\usepackage{xspace}
\usepackage{relsize}
\usepackage{booktabs}
\usepackage{tikz,siunitx,mwe}
\usepackage{subfig}
\usepackage{siunitx}
\usepackage{multirow}

\usepackage{graphbox}
\usepackage{cleveref}
\usepackage[super]{nth}

\def\code#1{{{\relsize{-1}\texttt{#1}}}\xspace}

\usepackage{color}

\newcommand{\textembeddedimg}[6]{
  \begin{tikzpicture}
    \draw node[name=micrograph] {\includegraphics[width=#2]{#1}}; 
    \draw (micrograph.north west)node[anchor=north west,yshift=-4.5, xshift=3,#6]{\LARGE{(#5)}};
  \end{tikzpicture}
}

\usepackage{xcolor}
\definecolor{dkgreen}{rgb}{0,0.6,0}
\definecolor{dkblue}{rgb}{0,0,0.6}
\definecolor{gray}{rgb}{0.5,0.5,0.5}
\definecolor{mauve}{rgb}{0.58,0,0.82}
\definecolor{commentgreen}{RGB}{2,112,10}
\definecolor{eminence}{RGB}{108,48,130}
\definecolor{weborange}{RGB}{255,165,0}
\definecolor{frenchplum}{RGB}{129,20,83}
\definecolor{darkmagenta}{RGB}{139, 0, 139}
\usepackage{listings}
\title{Data Parallel Rendering of Large Unstructured Data with Natively Non-Convex Partitions}

\title{Data Parallel Rendering of Non-Trivially Partitioned Unstructured Data using Shell Marching and Deep Compositing}

\title{Data Parallel Rendering of Large Unstructured Data in the presence of Non-Trivial, Non-Convex Pre-Partitioning}

\title{Data Parallel Rendering of Large, Unstructured, and Non-Convexly Partitioned Data}

\title{GPU-based Data-parallel Rendering of Large, Unstructured, and Non-convexly Partitioned Data}

\author{Alper Sahistan\thanks{equal contribution} \thanks{alper.sahistan@bilkent.edu.tr} \\
\scriptsize Bilkent University \\
\and Serkan Demirci\footnotemark[1] \thanks{serkan.demirci@bilkent.edu.tr} \\
\scriptsize Bilkent University
\and Ingo Wald\thanks{iwald@nvidia.com} \\
\scriptsize NVIDIA Corporation
\and Stefan Zellmann \thanks{stefan.zellmann@h-brs.de} \\
 \scriptsize Bonn-Rhein-Sieg University of Applied Sciences
\and Jo\~{a}o Barbosa\thanks{jbarbosa@macc.fccn.pt} \\
\scriptsize INESC-TEC \& University of Minho
\and Nathan Morrical\thanks{natemorrical@gmail.com} \\
 \scriptsize University of Utah
\and U\u{g}ur G\"{u}d\"{u}kbay\thanks{gudukbay@cs.bilkent.edu.tr} \\
\scriptsize Bilkent University}

\teaser{
  \centering
  \resizebox{0.99\textwidth}{!}{
    \setlength\tabcolsep{0pt}
    \includegraphics[height=5.3cm]{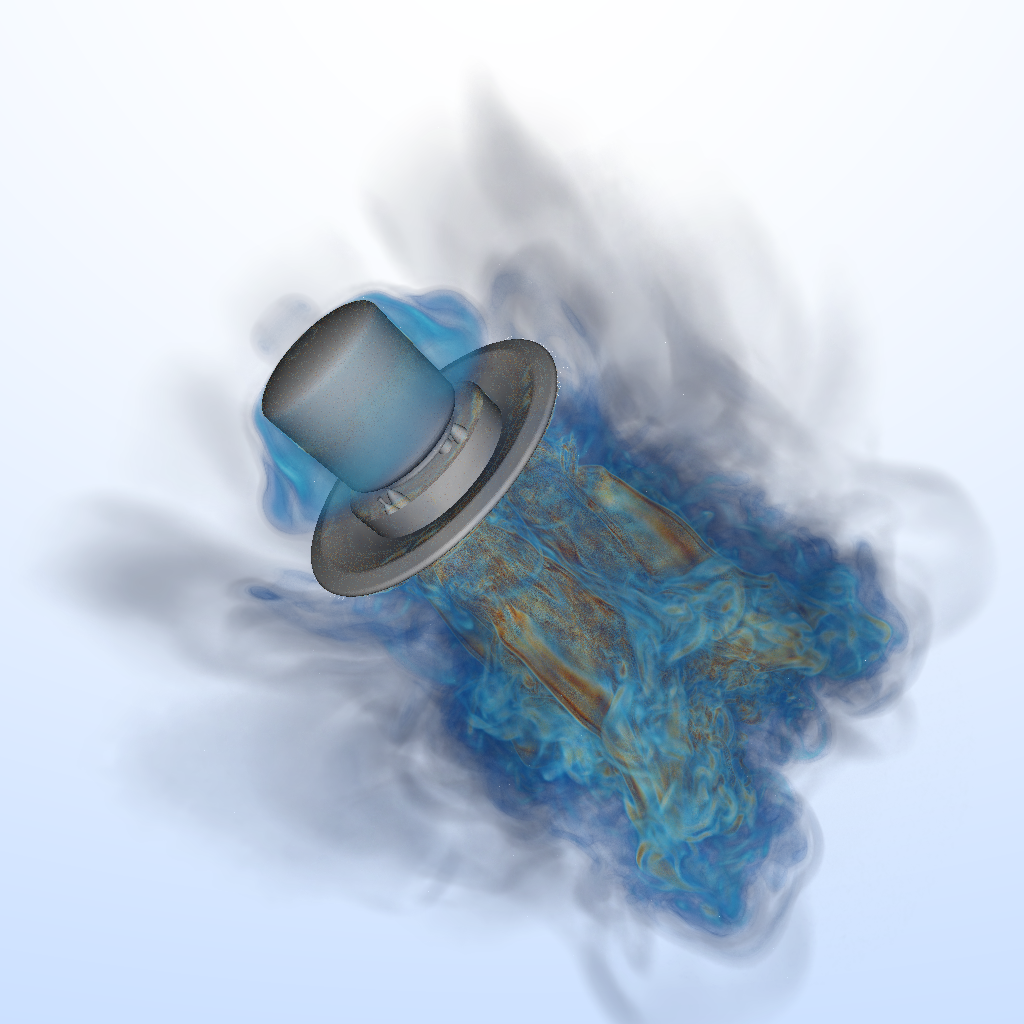}
    \includegraphics[height=5.3cm]{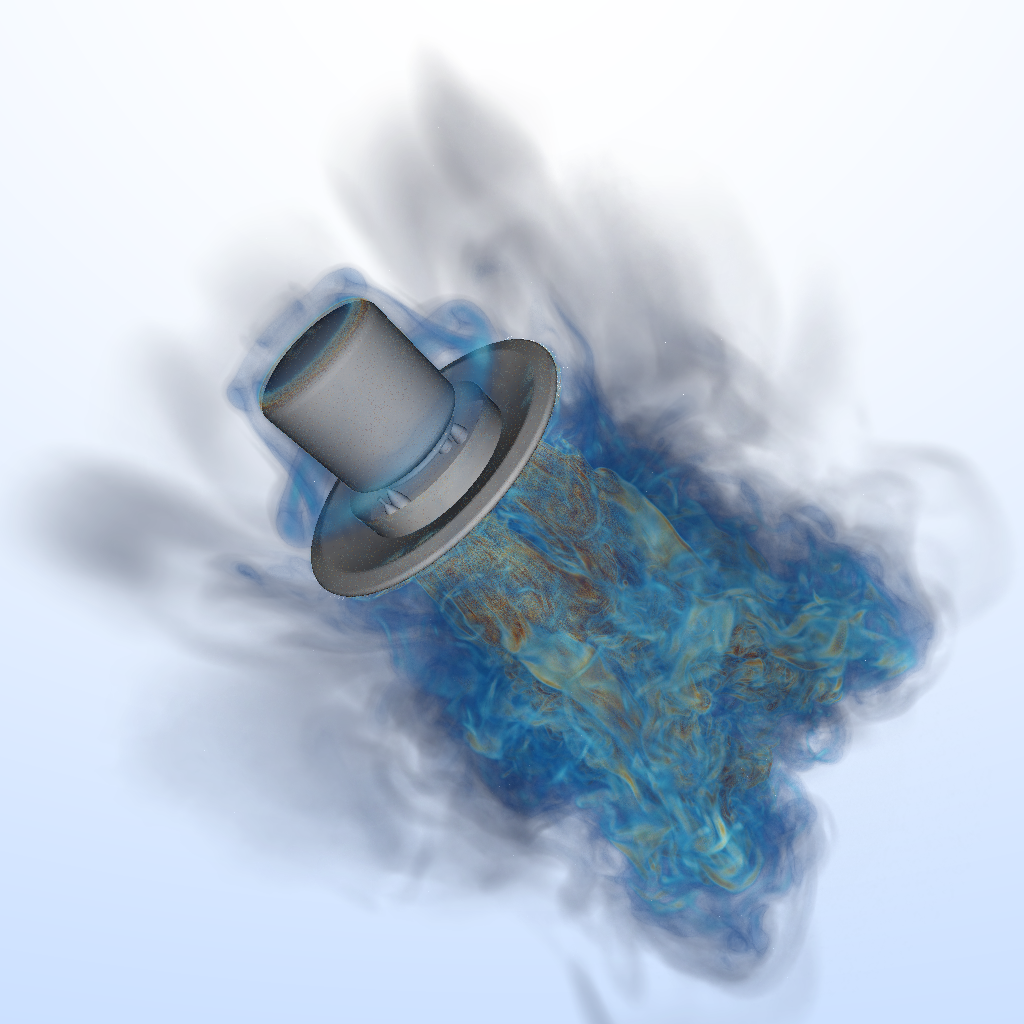}
    \includegraphics[height=5.3cm]{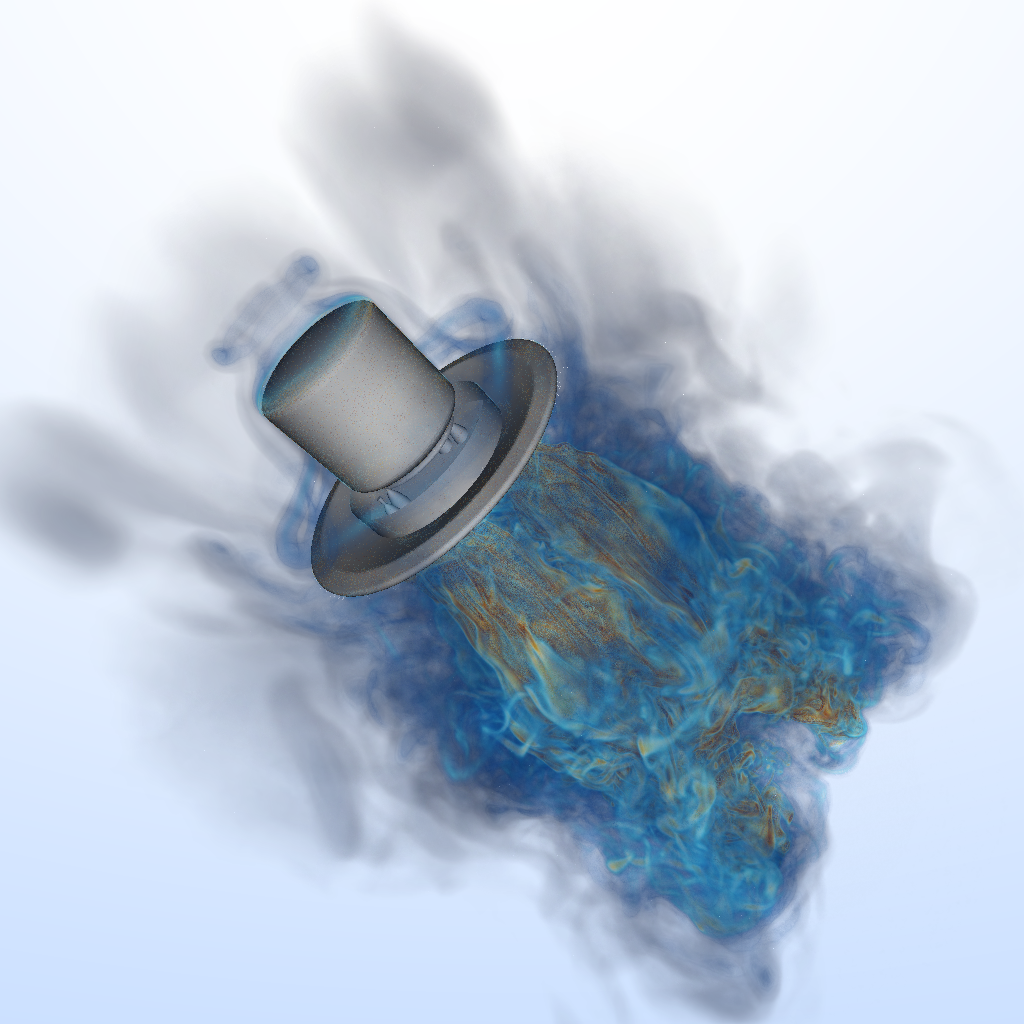}
    \includegraphics[height=5.3cm]{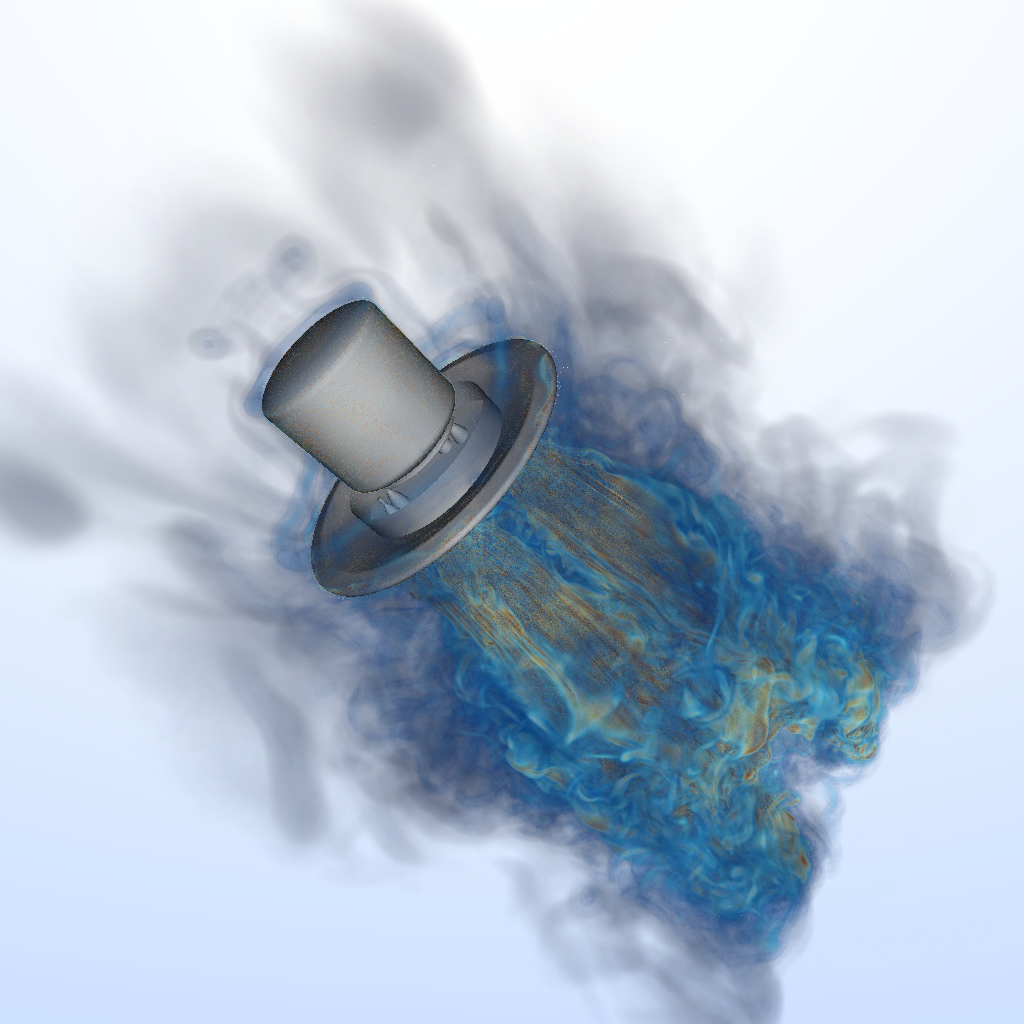}
    \includegraphics[height=5.3cm]{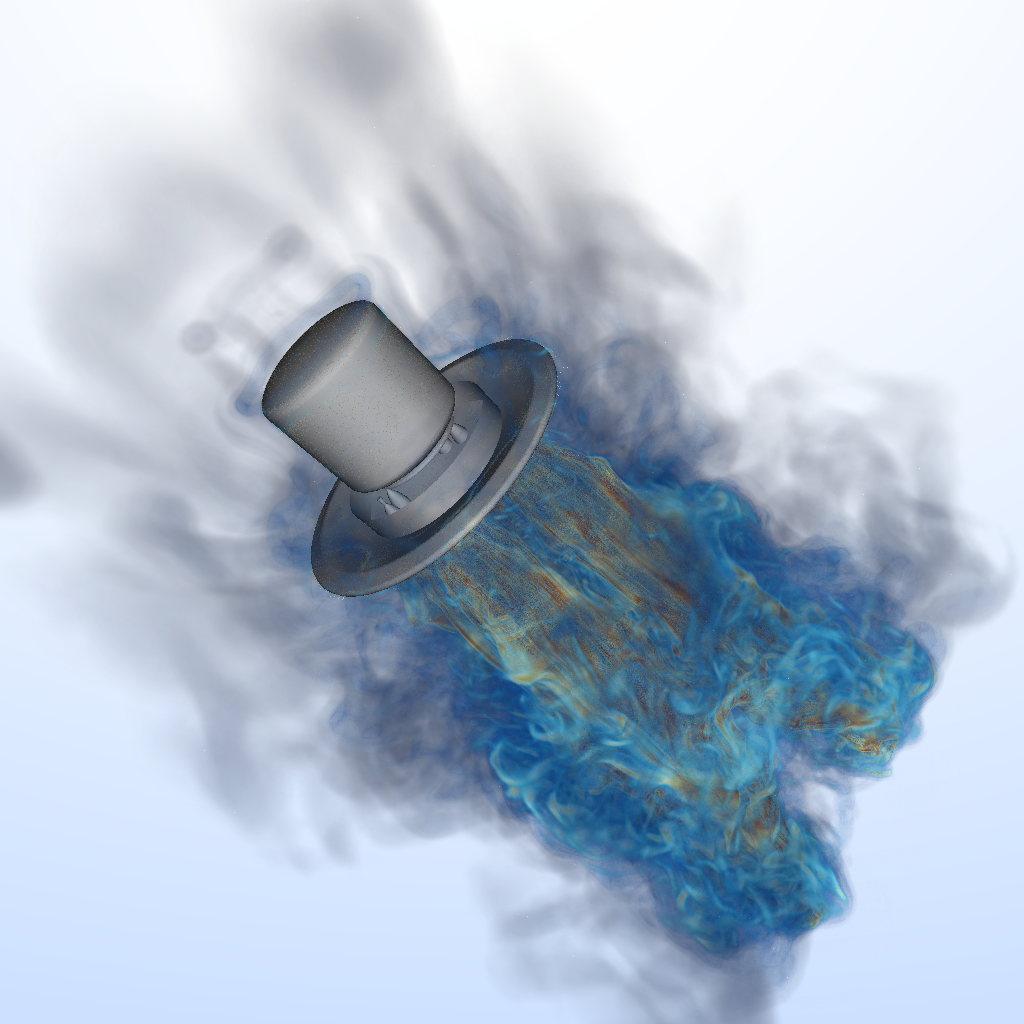}
  }
  \caption{Data-parallel rendering of the ``Small Mars Lander'' computational fluid dynamic data set, which is comprised of 72 clusters of finite elements. The images are rendered on 18 compute nodes and a total of 72 NVIDIA Quadro RTX5000 GPUs of the Frontera RTX partition at the Texas Advanced Computing Center (TACC). The image resolution is $1024\times1024$. Frame rates of these images are 15.14, 13.63, 15.04, 15.00, and 14.99 frames per second (fps), respectively.}
  \label{fig:teaser}
}

\abstract{Computational fluid dynamic simulations often produce large clusters of finite elements with non-trivial, non-convex boundaries and uneven distributions among compute nodes, posing challenges to compositing during interactive volume rendering. Correct, in-place visualization of such clusters becomes difficult because viewing rays straddle domain boundaries across multiple compute nodes. We propose a GPU-based, scalable, memory-efficient direct volume visualization framework suitable for in~situ and post~hoc usage. Our approach reduces memory usage of the unstructured volume elements by leveraging an exclusive or-based index reduction scheme and provides fast ray-marching-based traversal without requiring large external data structures built over the elements themselves. Moreover, we present a GPU-optimized deep compositing scheme that allows correct order compositing of intermediate color values accumulated across different ranks that works even for non-convex clusters. Our method scales well on large data-parallel systems and achieves interactive frame rates during visualization. We can interactively render both Fun3D Small Mars Lander (14 GB / 798.4 million finite elements) and Huge Mars Lander (111.57 GB / 6.4 billion finite elements) data sets at 14 and 10 frames per second using 72 and 80 GPUs, respectively, on TACC's Frontera supercomputer. 
}


\CCScatlist{ 
  \CCScat{Computer Graphics}{Three-Dimensional Graphics and Realism}{Raytracing}
  \CCScat{Human-centered computing}{Visualization}{Visualization systems and tools}
  \CCScat{Human-centered computing}{Visualization}{Visualization application domains}{Scientific visualization}
}

\begin{document}

\firstsection{Introduction}
\maketitle

Compute power has increased tremendously over the years, which has led to an increase in the number of operations performed by supercomputers, both simultaneously and on single processors. At the same time, memory bandwidth has also increased, but not at the same rate as computing power. This tendency in computer architectures and their development also affected simulation codes, which focus the computation on regions of space where the data contains high-frequency features to make good use of the available bandwidth. One way of doing this is to employ generally unstructured mesh topologies comprised of finite elements.

A similar gap between computing and memory sizes is also observed. Supercomputers must use their \emph{combined} memory resources to fully utilize available compute nodes and accommodate data domains of extensive simulations and hence require data-parallel processing to make optimal use of large, distributed memory systems. Because the data is so big and compute resources are limited, simulation codes often support in~situ visualization and steering and use the supercomputer's processors to visualize the data. In this way, users can explore and potentially influence intermediate outcomes of a simulation interactively.

These circumstances, and particularly their combination, make scientific visualization extremely challenging. As we usually cannot redistribute data for postprocessing while the simulation is running, we have to work with the distribution of finite elements given by the simulation and its assignment to the processors, which was optimized for the simulation, is not necessarily well-suited for data-parallel rendering. 

Our work is motivated by the Fun3D Mars Lander simulation data set containing a collection of massive retropropulsion simulations with mixed elements and various per-vertex scalar fields. The scalar fields come as an animation sequence with multiple timesteps. While the scalar fields are animated, the mesh topology (vertex position and connectivity) remains stationary over time, which is typical for such kinds of simulations~\cite{Anderson99,Fun3DManual,moran2020fun3d}. Fig.~\ref{fig:teaser} shows an animation frame sequence obtained from the smaller of the two Mars Lander data sets.

We aim to use the exact data distribution (geometry and clusters) generated by the Fun3D system as provided~\cite{moran2020fun3d} by NASA. Since such Computational Fluid Dynamics (CFD) simulations produce non-trivially partitioned and non-convex data, they pose significant challenges for visualization algorithms. \autoref{tab:fun3d} provides data set statistics of two such data sets, i.e., the small and huge Mars Lander. Our approach is not \emph{directly} in~situ but rather emulates such a scenario, and if the simulation code were adapted accordingly, it would be directly applicable to in~situ strategies, simulation steering, and post~hoc analysis.

\begin{table}[htbp]
  \caption{The Fun3D ``Mars Lander'' data set statistics. These data sets do not contain hexahedral elements.}
  \centering{\small
  \begin{tabular}{@{}lc@{\hskip8pt}c@{\hskip8pt}c@{\hskip8pt}c@{\hskip8pt}c@{\hskip8pt}c@{}}
      \toprule
      & \multicolumn{5}{c}{Element counts} & ~ \\
      \cmidrule(rr){3-5}
      Model & Vertices & Tetrahedra & Pyramids & Wedges & Clusters & Size (GB)\\
      \midrule
      Small & 145 M & 766 M & 47.5 K & 32 M  & 72 & 14\\
      Huge & 1.2 G & 6.12 G & 285 K & 256 M & 552 & 112\\
      \bottomrule
      \end{tabular}
  }
\label{tab:fun3d}
\end{table}

The data-parallel partitions of finite elements, which we call \emph{clusters}, are typically non-convex, also  clusters from different compute nodes touch at the boundaries. This has numerous implications for scientific visualization algorithms and their particular implementation. Many out-of-the-box standard \emph{software} solutions for DVR fall short when handling such clusters~\cite{garrity1990rtirregvolume, weiler2003hardwarebased}. On top of that, data-parallel rendering usually employs a compositing scheme that works best with convex clusters. Another alternative to handle large-scale data can be compression-like big mesh compaction~\cite{wald2022bigmesh}; even then, applications might use a data-parallel paradigm due to memory and bandwidth limitations. However, long build times and extra complexity introduced with multiple scalar fields/timesteps make it an unviable option for our purposes. Besides, in an in~situ scenario, there may not be enough memory and computation resources available to achieve that sort of data wrangling.

Ray-tracing and its derivatives are among the alternatives to rendering volumetric data sets directly. However, with data-parallel applications, rays that straddle the domain boundaries of different clusters would consequently have to integrate the data across multiple compute nodes, which, if na\"{\i}vely implemented, would result in non-trivial communication patterns. A critical point here is correct alpha compositing, which requires the variable number of partially composited \emph{fragments} to be processed in visibility order. Particularly, typical image compositing algorithms that require the data domains to be convex and are, e.g., implemented in the popular IceT  library~\cite{icet}, are unsuitable for this.

We propose a scalable distributed direct volume visualization approach to address these challenges. We use a ray-marching algorithm that can traverse unstructured meshes containing various primitives, including tetrahedra, pyramids, wedges, and hexahedra. The algorithm exploits XOR-based index reduction to decrease information transfer and storage for tetrahedra, pyramids, and wedges. We support this ray-marcher with a \emph{shell-to-shell traversal} technique that can find each ray's entry and exit points for convex and non-convex boundary geometries across multiple MPI nodes~\cite{sahistan2021Shell}. We also present a \emph{deep compositing} scheme that can merge fragments from convex and non-convex clusters across compute-nodes. Specifically our contributions are as follows:
\begin{itemize}
\item a memory-efficient, low-overhead ray-marching strategy that can handle unstructured volumetric meshes with different types of primitives, including tetrahedra, pyramids, hexahedra, and wedges;
\item a shell-to-shell traversal scheme and a deep compositor supporting convex and non-convex geometry with proper depth ordering;
\item a GPU-optimized, scalable method that uses native data partitioning and allows data-parallel rendering and animation of massive (time-varying) data sets with small support structures at interactive rates.
\end{itemize}

The remainder of the paper is organized as follows. Section~\ref{sec:related} provides related work on direct volume rendering of unstructured meshes, data-parallel rendering, and in~situ visualization. Section~\ref{sec:problem} gives problem statement. Section~\ref{sec:method} describes our method and details its steps, which are data preparation (\autoref{sec:data-preparation}), per-node segment generation using shell-to-shell traversal~(\autoref{sec:shell-to-shell}), per-segment volume integration process (\autoref{sec:integrate-segments}). Section~\ref{sec:evaluation} presents experimental results, focusing on the scalability of our approach in terms of computational cost and memory consumption. Section~\ref{sec:discussion} discusses performance and memory scalability, especially regarding our test scenarios. Section~\ref{sec:conclusion} concludes and provides further research areas. 

\section{Related Work}
\label{sec:related}

We give related works regarding direct volume rendering (DVR) of unstructured meshes, data-parallel rendering, and in~situ visualization.

\subsection{Unstructured Volume Rendering}

There are notable works proposed to render unstructured finite element meshes~\cite{nelson2006isosurface, Vo2007irun, marmitt2008cpuvolume, muigg2011interactivevol}. Two predominant strategies to render unstructured volumes are \emph{point-query sampling}, e.g.,~\cite{morrical2020rtxpointlocext} and \emph{ray-marching}~\cite{shirley1990rastertet}.

In point-query sampling, zero-length rays are used to probe into an acceleration data structure, for example, a \emph{bounding volume hierarchy} (BVH), that was built over the elements to sample the volumetric data adaptively. These structures are traditionally used to find ray-particle collisions per frame for all pixels. Rathke et al.~\cite{rathke2015simd} propose a min/max BVH that speeds up the element look-up processes for samples and iso-surfaces. Wald et al.~\cite{wald2019rtxpointloc} use point location queries on tetrahedral meshes by utilizing NVIDIA's ray-tracing (RT) cores. This work is later extended by Morrical et al.~\cite{morrical2020rtxpointlocext} to include all unstructured elements. Due to the low number of samples taken per ray, these approaches can produce fast but noisy results, thus requiring many samples to be taken over time for a converged image. To further accelerate the process of convergence and sampling, empty space skipping~\cite{kruger2003accelerationgpu, hadwiger2018sparseleap} or adaptive sampling strategies~\cite{Wang2020noveladaptivesampling,szirmaykalos2011freeps} can be leveraged. The RTX hardware also can be exploited for empty space skipping and adaptive sampling~\cite{morrical2019spaceskip}. 

Standard ray-marching accumulates many samples while tracing rays without using external acceleration structures~\cite{marmitt2006fastraytet, weiler2003hardwarebased}. Usually, \emph{marching} is performed via visibility sorting or element connectivity. A well-known way to render tetrahedral meshes without connectivity information is by Shirley and Tuchmann~\cite{shirley1990rastertet}. Since visibility sorting tends to be very costly, several researchers turned their attention to connectivity storage, eliminating the need for sorting. Aman et al.~\cite{aman2021bth, aman2021compact} introduce a tetrahedra traversal algorithm that optimizes intersection tests by using 2D projection while still maintaining a connectivity list. However, most of these works are limited to pure tetrahedral meshes. Muigg et al.~\cite{muigg2011interactivevol} propose a ray-marching algorithm that can handle non-tetrahedral elements and non-convex bounding geometry by storing compact face-based connectivity lists and projecting vertices to a ray-centric coordinate system for intersections. When doing ray-marching, one also needs to find the first element where the ray first enters the volume; \cite{sahistan2021Shell} showed how this can be done with RTX hardware by building a BVH over the shell and tracing rays.

Ray-marching techniques that utilize connectivity are particularly appealing to our purposes because many modern simulation systems already store connectivity data. Therefore they can be used to avoid worsening high memory pressure situations. 

\subsection{Data-parallel Rendering}

Parallel approaches should be exploited to visualize massive simulation data sets with proper timings. There are various means to partition the workload and data among many compute nodes. Distributing data pieces (clusters) between nodes (i.e., data-parallel rendering or \emph{sort-last}) is a popular method employed by recent works~\cite{larsen2015raytracingdataparallel, castanie2006distributedsharedmemory}. Some works also propose an image-order partitioning (\emph{sort-first}) where work is distributed over pixel regions~\cite{brownlee2013imageparallel,biedert2018hwacceleratedmultitile}. There are also hybrid approaches~\cite{biedert2017taskbasedparallel, cao2019parallelvis}, which aim to address load-balancing issues by leveraging both perspectives.

Sort-last algorithms allow for a static geometry assignment at the cost of exchanging intermediate images. Because of the static geometry assignment, sort-last is the most popular rendering algorithm on distributed memory systems. However, correctly and efficiently compositing intermediate images that generally overlap is challenging. Image-based compositors, such as IceT\cite{icet, moreland2011icet}, produce a single intermediate image per node, which is not suitable for clusters with non-convex domains. One of the recent works that tackle the sort-last compositing problem is by Grosset et al.~\cite{grosset2016imagecompositing}, which reduces delays and communications by implementing a spatiotemporally-aware compositor. Their approach uses ``chains'' that determine the blending order of each strip of the image. Usher et al.~\cite{usher2019distributedfb} introduce \textit{Distributed FrameBuffers}, a method that breaks the image processing operations into tiles of ranks with independent dependency trees. The Galaxy framework~\cite{abram2018galaxy} displays the idea of an asynchronous frame buffer, which leverages independent pixel updates sent from a server while allowing incremental refinements to the final image over time.

Many recent works are proposed for optimizing workloads minimizing communication costs while generating images with the highest possible accuracy. Ma~\cite{liukwan1995parallelvolumedistributed} introduces a data-parallel unstructured volume rendering method with the ability to handle non-convex data boundaries properly. Similar to our shells (see \autoref{sec:shell-to-shell}), their technique makes use of a \emph{hierarchical data structure} that allows accessing the boundary faces and ray-casting operations from these faces. This work also describes how to do compositing in the correct order. However, unlike our deep compositing (\autoref{sec:deep-compositing}), they prefer sending smaller many messages between compute nodes during rendering. Some of these ideas are later extended to utilize asynchronous load balancing via object and image-order techniques~\cite{liukwan1997scalableparallelcell}. However, this work uses cell-projection techniques rather than ray-casting.

Childs et al.~\cite{childs2006hybridmassive} layout a two-stage framework that first samples a  $m \times n\times k$ view-aligned grid ---where m and n denote the pixel resolution and $k$ is the sample per pixel--- then composites these samples in the proper viewing order. In the sampling stage, first, they sample what they consider to be small-sized elements. Then, they distribute the large elements to processors, where they are sampled to balance the load. This work is later extended by Binyabib et al.~\cite{binyabib2019hybrid} by proposing a many-core hybrid scheme where they employ sampling over a similar view-aligned grid, but this scheme allows successive $k$ samples in the same pixel and node to be partially composited, reducing the memory footprint. Our deep compositing algorithm is an extension of the algorithms by Childs and Binyabib et al. in the regard that ours can also handle jagged cluster boundaries. However, the view-aligned grid-based sampling is infeasible for our purposes as, in theory, it will waste precious memory resources for large framebuffers.

In theory, the 3D rasterization process required by Childs et al. will also be sensitive to overdraw when millions of elements fall within the same grid cell. Finally, both of these works' image-order load balancing method requires large elements to be either replicated or moved to some other nodes, thus requesting additional memory, which may not always be present given an in~situ scenario. We also acknowledge GPU architectures improve with new divergence handling methods and ray-tracing (RT) cores, so the adaptations we propose in this work are necessary.

Unlike these prior works, our method is tailored for modern GPUs, which minimizes the costs of compositing and sorting operations. We also do not have to buffer every sample along the pixels since we ray-march through each segment to determine partial samples. Finally, unlike these works, our approach does not require re-distributing or replicating elements across nodes to render the data.

\subsection{In~situ Visualization}

File I/O has long been a bottleneck of high-performance computing. To overcome this hindrance, in~situ visualization couples computation and visualization together, thus, enabling the users to tap into a running simulation. In~situ visualization has many merits, such as examining the data, doing numerical queries, and generating graphical outputs while the simulation executes. Moreover, it allows verification so that the simulation may be stopped or modified, saving both time and computation resources~\cite{childs2020terminsitu}. We find our approach in line with in~situ applications because of our ability to generate correct images with little to no support from additional acceleration structures at interactive rates.

There are notable in~situ applications used in industry~\cite{ayachit2015catalyst, kuhlen2011libsim} that render outputs generated by infrastructures like Strawman~\cite{Larsen2015strawman} or Ascent~\cite{larsen2017alpine}. In addition to standard systems, various recent algorithmic novelties have been proposed to handle time-varying data generated by the simulations. Yamoka et al.~\cite{yamaoka2019adaptivetimestepinsitu} illustrate a method that adapts the timestep sampling rate according to variations in the probability distribution function (PDF) estimation of the connected simulation. Aupy et al.~\cite{aupy2019highthroughputinsitu} give a model that allows them to analyze simulations, and then they use this model to formulate high-throughput scheduling. DeMarle and Bauer~\cite{demarle2021temporalinsitu} propose a temporal cache scheme that keeps much time-varying information produced from a running simulation, which can later be stored according to a pre-defined trigger. Marsaglia et al.~\cite{marsaglia2018explorativevis} introduce an error-bound in~situ compression scheme that allows saving complete spatiotemporal simulation data. Our proposed method only requires a couple of lightweight structures alongside what is already being kept in simulations. Furthermore, observing recent trends from these approaches, we see no major issues that stop our method from being used alongside current in~situ systems.

\section{Problem Statement}
\label{sec:problem}

\begin{figure*}
    \centering
  \resizebox{0.99\textwidth}{!}{
    \textembeddedimg{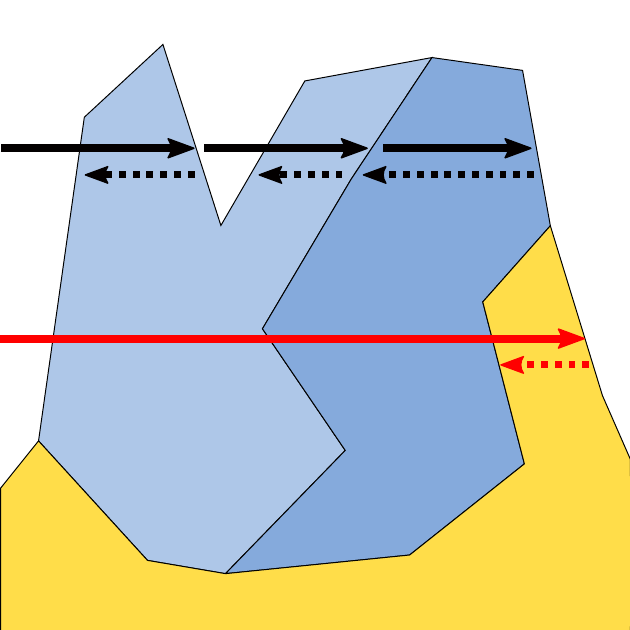}{0.24\textwidth}{}{}{a}{black}\hfill
    \textembeddedimg{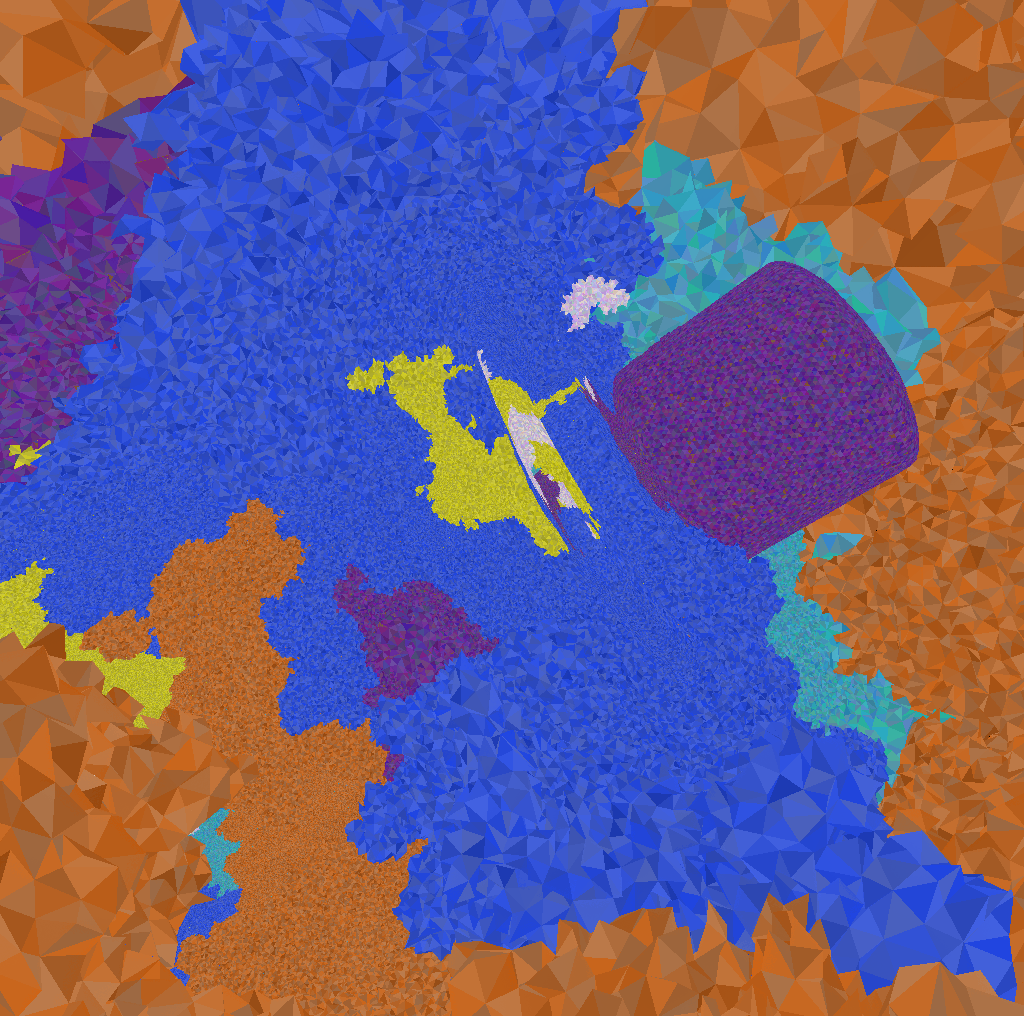}{0.24\textwidth}{}{}{b}{white}\hfill
    \textembeddedimg{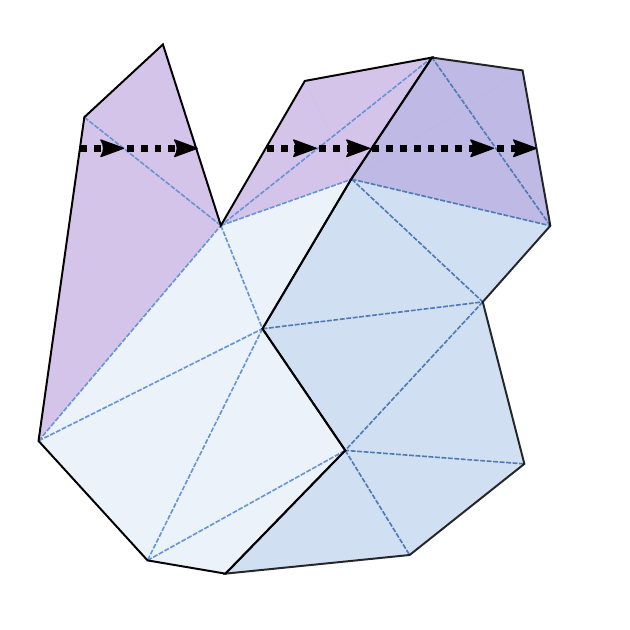}{0.24\textwidth}{}{}{c}{black}\hfill
    \textembeddedimg{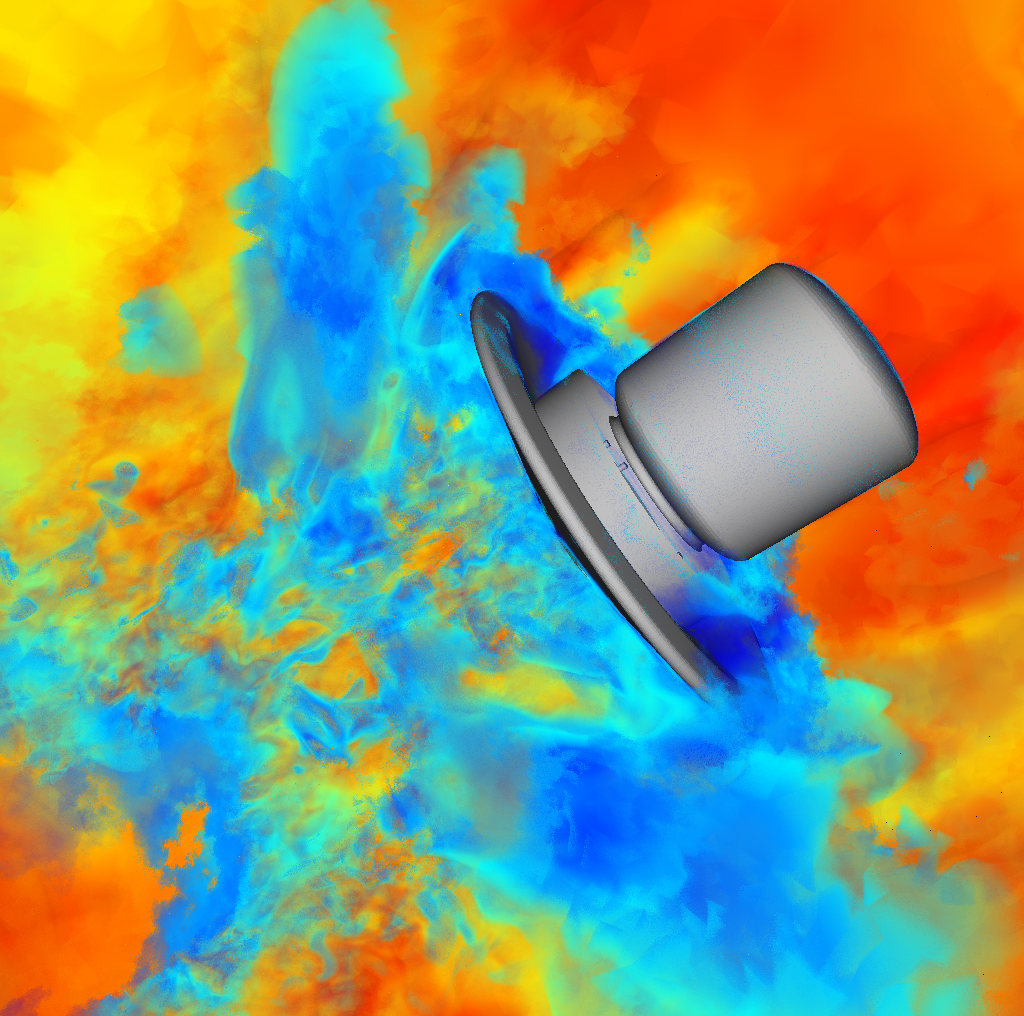}{0.24\textwidth}{}{}{d}{black}
  }
  \caption{Overview of the volume integration process. (a) Shell-to-shell traversal for non-convex elements: rays with front-face culling are sent to find exit faces, and backward rays are cast from exit faces to find the entry faces over the shells. The blue and light blue clusters lie in the same MPI rank, but the yellow cluster is in another rank; hence, the shells of the two blue clusters are traced in order, but the yellow cluster's shells get traced in parallel. (b) The shells of the first 46 clusters of Small Lander. Each base color represents a different cluster's shell. (c) The ray-marching process is illustrated for the two blue clusters given in (a). (d) DVR of the same subset given in (b).
  }
\label{fig:overview}
\end{figure*}

Modern simulation data is becoming more extensive and complex each day. With the unstructured volume data sets, like The Fun3D Mars Lander that contains many parts (i.e., \emph{clusters}) with non-convex boundaries (see \autoref{fig:overview} (b)), robust data-parallel solutions are needed. Moreover, the generation of such data sets require carefully tuned simulations. In~situ visualization can be utilized to verify the correctness of simulations by allowing visuals to be taken in simulation-time. However, due to the different requirements of the simulation and visualization algorithms, the volume rendering at interactive rates can be challenging. One major problem of such simulations is that the data distribution is generally unbalanced for visualization purposes. Due to time and memory costs, re-distributing the data does not offer a feasible option. Besides, allocating solely visualization-related acceleration structures over all elements may not be possible since nodes may not have enough space. 

The data-parallel rendering requires each partition to be on a separate computer node, where each node renders a portion of the final image. These portions are called \emph{fragments}, and in our approach, they are generated per \emph{ray segment}. Ray segments are defined between an entry and exit position of a \emph{shell} (faces defined by cluster boundaries), so for non-convex cluster boundaries, there might be more than one fragment since there might be more than one ray segment. Furthermore, these non-convex shells can be on different nodes and interleave each other, which makes the correct order compositing extremely difficult.

\section{Proposed Approach}
\label{sec:method}

We tackle the given problem by introducing a framework that supports interactive visualization of large, unstructured, and non-convexly partitioned volumetric data sets, animation of fixed topology data sets, and compatibility with the given data partitioning and the number of nodes in the simulations; i.e., native data, without re-partitioning or simplifications, and proper compositing even in the presence of non-convex data. Assuming that the data is natively pre-partitioned and distributed to different ranks, the proposed approach for rendering consists of four steps:

\begin{enumerate}
\item Each node generates connectivity information, shell-BVH, and XOR-compacted geometry representation required by our ray-marcher (\autoref{sec:data-preparation}).
\item Rendering starts at each node by tracing two rays through each shell to create segments (\autoref{sec:shell-to-shell}). \autoref{fig:overview}~(a) illustrates the shell-to-shell traversal.
\item Each node performs volume integration (cf.~\autoref{fig:overview}~(c)) via ray-marching, creating one RGBA-Z tuple; i.e., fragment per each segment, resulting in potentially multiple fragments for each pixel (\autoref{sec:integrate-segments}). \autoref{fig:overview} (d) depicts an integrated volume output using the shells from \autoref{fig:overview} (b).

\item Finally, we apply a GPU-optimized ``deep compositing'' technique in which different ranks exchange their respective fragments and composite them in the proper order (\autoref{sec:deep-compositing}).
\end{enumerate}

In an offline rendering mode, each step is executed once in the given order; however, the last three steps are repeated repeatedly under interactive exploration. Besides, the first step's results can be cached for future use.

\subsection{Data Preparation}
\label{sec:data-preparation}

We describe the connectivity information generation, shell-BVH construction and XOR-compacted geometry representation creation required by our ray-marcher.

\subsubsection{Connectivity Generation}
\label{sec:connectivity-gen}

Our method needs to know the neighboring elements' indices to perform element marching. We generate the connectivity information by matching the element faces in the preprocessing step. We separate vertex and connectivity information and store the neighbor indices in an external buffer to keep the elements and neighbor indices aligned in memory. Although we picked this way of processing connectivity it should be noted that this buffer can be in any shape or form as long as one can access the next element from the current element using a face that is shared by both. Thus this part can be adapted to fit simulation or application's needs.

\subsubsection{Per-Node Shell Generation}
\label{sec:per-node-shell-gen}

Our approach handles volumetric data that may contain convex and non-convex clusters. To this end, we identify boundary geometry for each cluster present per node by looking at elements which are missing a neighbour from the connectivity generation step. This boundary geometry comes in the form of triangles and quads, which we call \emph{shell-faces}. 

We utilize a method similar to the one described in Sahistan et al.~\cite{sahistan2021Shell}, where we identify each shell-face using connectivity. We mark the faces from elements with missing neighbors as shell-faces. We use triangles as provided and triangulate quadrilateral elements. We keep a list of triangle indices stored along the shell-BVH. We reserve four indices for each shell triangle: the first three are triangle indices, and the last one points to the volume element behind that triangle. The lower two bits of the fourth index signifies the element type (i.e., tetrahedron, pyramid, wedge, or hexahedron), and the remaining 30 bits is an index into the list of elements; this index is required to start marching. This encoding is similar to the BVH-node memory layout used by PBRT~\cite{pharr2021pbrtaccel}. We build our shell-BVH using OptiX~\cite{optix, wald2020owl} to exploit NVIDIA GPU's RTX cores for hardware-accelerated shell-to-shell traversal.

\subsubsection{XOR-compaction}
\label{sec:xor-comp}

\begin{figure}[htbp]
\begin{minipage}[t]{.22\columnwidth}
{\small
\begin{lstlisting}[language=c++]
struct Tet{
  uint  vx;
};


\end{lstlisting}
}
\end{minipage}
\hfill 
\begin{minipage}[t]{.22\columnwidth}
{\small
\begin{lstlisting}[language=c++]
struct Pyr{
  uint dx;
  uint diag[2];
  uint top;
};
\end{lstlisting}
}
\end{minipage}
\hfill
\begin{minipage}[t]{.22\columnwidth}
{\small
\begin{lstlisting}[language=c++]
struct Wed{
  uint dx[2];
  uint diag[2];
};

\end{lstlisting}
}
\end{minipage}
\hfill
\begin{minipage}[t]{.22\columnwidth}
{\small
\begin{lstlisting}[language=c++]
struct Hex{
  uint  v[8];
};


\end{lstlisting}
}
\end{minipage}

\begin{center}
\fbox{%
\includegraphics[width=0.28\columnwidth]{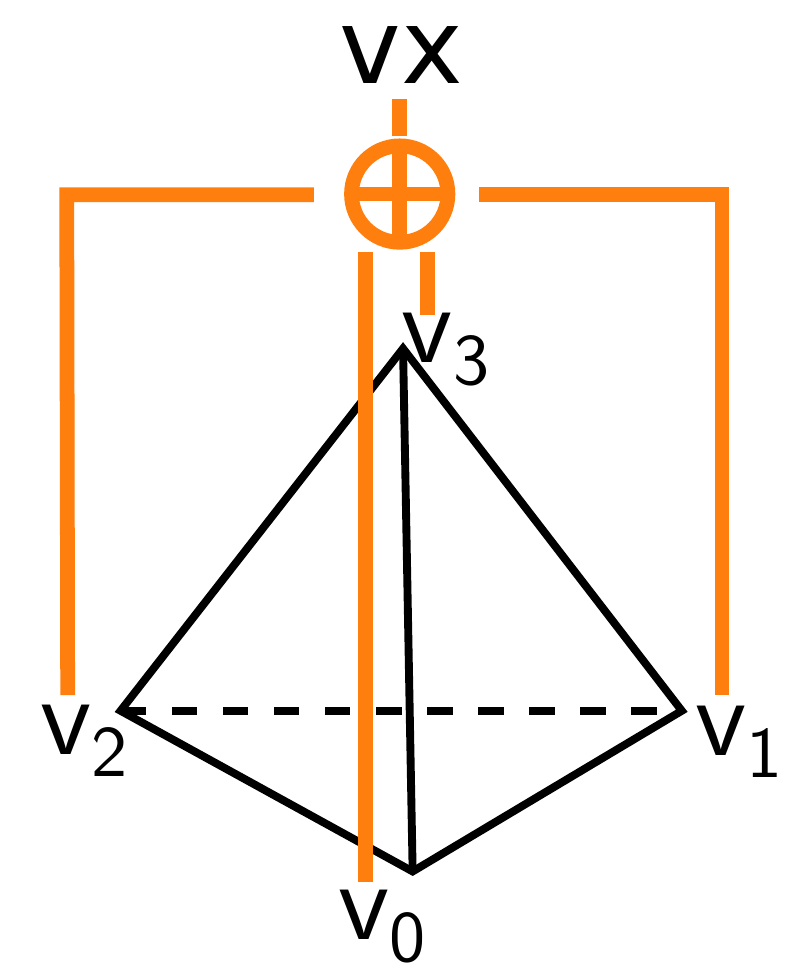}
}
\fbox{%
\includegraphics[width=0.28\columnwidth]{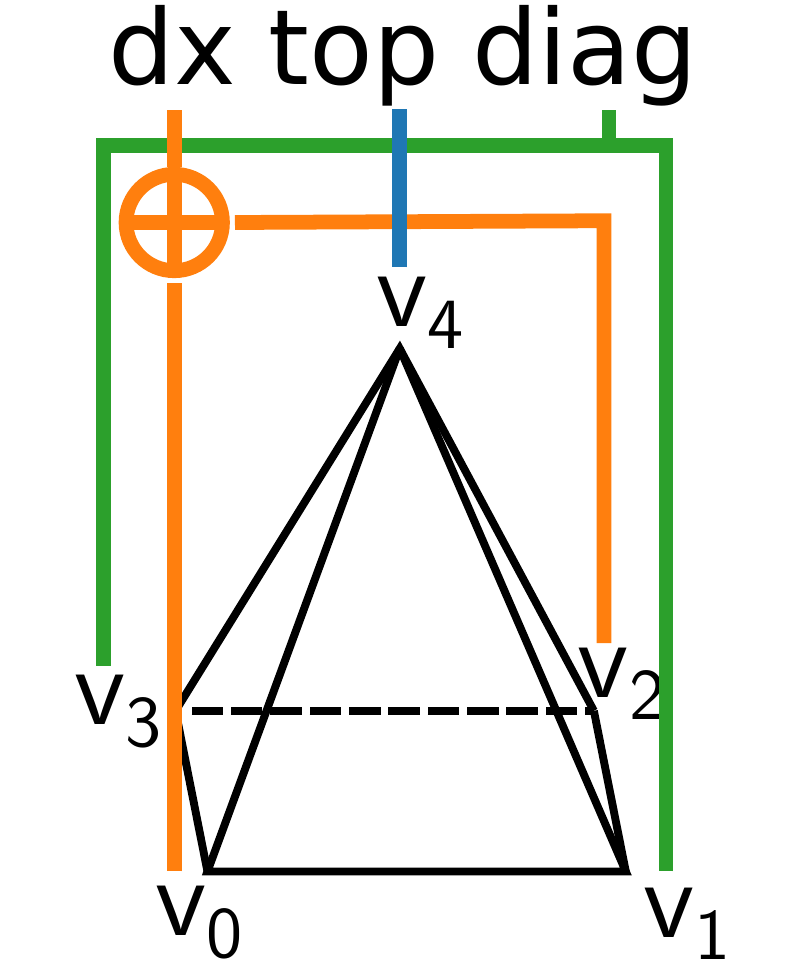}
}
\fbox{%
\includegraphics[width=0.28\columnwidth]{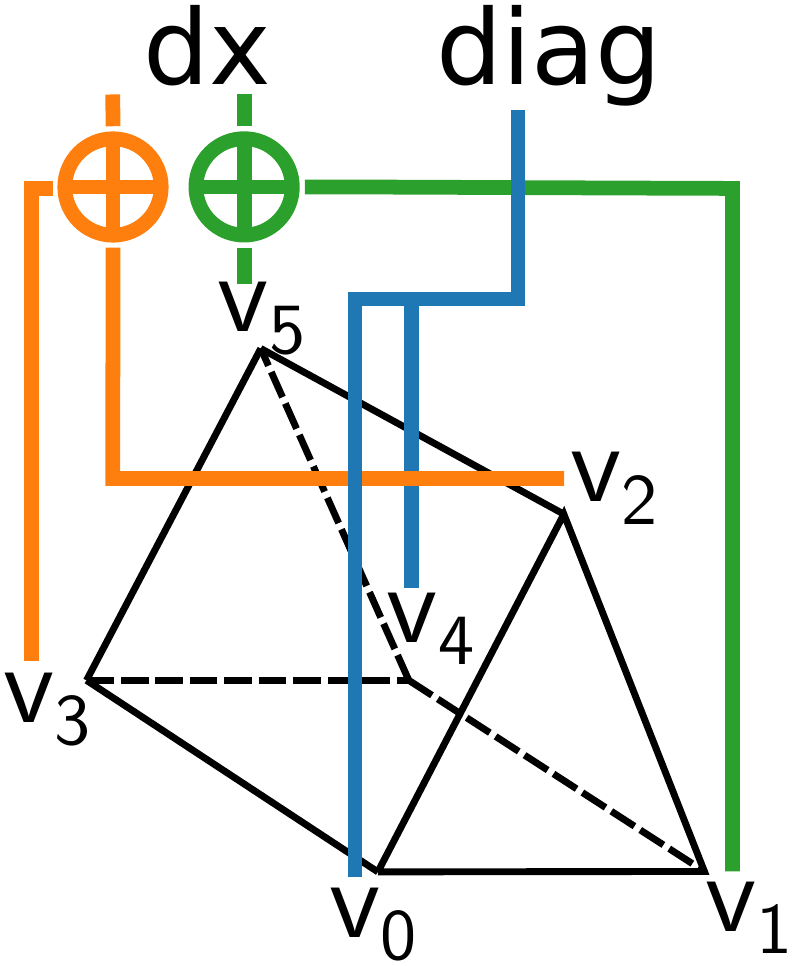}
}
\end{center}
\caption{XOR-compacted memory layouts (top) and geometric illustrations of XOR calculations for \code{Tet}, \code{Pyr}, and \code{Wed} (bottom). \code{Hex} does not have a compaction scheme. \texttt{uint} stands for unsigned integer. The ``$\oplus$'' symbol indicates the XOR operation. The total sizes of each struct are 4, 16, 16, and 32 bytes for Tet, Pyr, Wed, and Hex, respectively. The vertices are in VTK~\cite{VTK} ordering.}
\label{fig:memorylayouts} 
\end{figure}

Our compaction scheme exploits the following property of exclusive-or (XOR) operations: $(a \oplus b) \oplus a = b$. We can generalize this property to $n$ numbers if we know the XOR of $n-1$ numbers. Let $X$ denote the XOR of $n$ terms and $Y$ denote the XOR of any subset with $n-1$ terms. Then, we can simply XOR $X$ and $Y$ to find the remaining term. We can exploit this idea on connected volume elements in a ray's path during ray-marching. Since we know that some of the vertices are shared between neighboring elements, previously calculated XOR fields can be employed to reduce index information per element. This compaction requires the first face to be known to start ray-marching since all the other steps depend on the information obtained from the previous step. To handle this initial case, we use our shell faces that we store explicitly. After that, each step utilizes the march state to access the previous step's information. For each element except hexahedra, we store a different XOR-compacted structure, illustrated in \autoref{fig:memorylayouts}.  

We apply the tetrahedron compaction method by Aman et al.~\cite{aman2021bth}. A tetrahedron shares three of its four vertices with any other neighboring element. For this reason, a single XOR field is enough to construct the unshared vertex. Given the entry face that contains vertex indices $v_0, v_1, v_3$ and a compacted tetrahedron with $vx = v_0 \oplus v_1 \oplus v_2 \oplus v_3$ one can XOR all four of the integers to get the missing vertex index $v_2$.

For the 16 byte pyramid, we store one \code{dx} field that is the XOR of \nth{0} and \nth{2} vertex indices (according to VTK ordering), two vertex indices which happens to be the other diagonal of the quad (\nth{1} and \nth{3} vertices), and a top vertex index, which is always the \nth{4} vertex. During marching if the entry is from the quad face, the only missing index is the top vertex index, which is already explicitly stored. Otherwise, any triangle face should be composed of one of the explicitly stored diagonal vertices, the top vertex, and one vertex encoded in the \code{dx} field. Matching one of the vertices with one of the diagonal fields, we can decide which index to XOR with \code{dx}, thus obtaining one of the missing vertices. The other missing index for this case is the unmatched integer from \code{diag[2]}.

Our 16 byte wedge structure is composed of two \code{dx} and two \code{diag} fields. \code{dx} fields contain two XORs: the first one is the XOR of \nth{2} and \nth{3} vertex indices; the second one is the XOR of \nth{1} and \nth{5} vertex indices. \code{diag} explicitly stores \nth{0} and \nth{4} vertex indices. Like pyramids, ray-marcher's wedge construction has two high-level cases. If the entry is from a triangular face, any triangle face should be composed of one of the explicitly stored diagonal vertices and two vertices encoded in the different \code{dx} fields. By matching one of the diagonal vertex indices to one of the diagonal fields, we can construct two missing vertices from \code{dx} fields. The other missing index for this case is the unmatched integer from \code{diag[2]}. If the ray enters from a quadrilateral, it must contain one or both of the indices stored in \code{diag}. By matching diagonal vertices, we can determine the entry quadrilateral. Then, we have two cases. The first case where two of the entry quadrilateral's indices match both of the indices is shown in \code{diag}. We can get the two missing indices by XOR'ing \code{dx} fields with unmatched indices of the entry quadrilateral separately. In the second case (either one of the diagonals match with one of the quadrilateral face indices), we can immediately get one of the missing vertices from unmatched \code{diag}. Finally, we can use one of the \code{dx} fields to get the other missing vertex. 

Hexahedra have the minimum shared vertices ratio (0.5) among all element types and demand that four vertices be obtained on entry. It is challenging to find an XOR-based hexahedra compaction scheme that reduces to the closest alignment of 16 bytes. Therefore, we store all hexahedra indices without compaction according to the VTK mesh ordering.

\subsection{Per-Node Segment Generation via Shell Traversal}
\label{sec:shell-to-shell}

For each given shell-BVH per node, we initiate a step called \emph{shell-to-shell} traversal. This process is done for every ray per cluster and is fundamentally similar to~\cite{sahistan2021Shell}, which finds a \emph{segment} between entry and exit faces. \autoref{fig:overview} (a) illustrates this process, and the steps are as follows:
\begin{enumerate}
    \item Trace the ray through the shell-BVH with front-face culling from the ray origin. 
    \item If a ray hits a shell face, we then mark that face as the exit face and create a backward ray with the origin at the hit position.
    \item This backward ray is again traced using front-face culling to find an entry face.
    \item The found entry face contains four index values (\autoref{sec:per-node-shell-gen}), and the last one encodes the ID and the type of the element from where we start our ray-marcher (\autoref{sec:integrate-segments}).
\end{enumerate}

Some real-life data sets might have degenerate volume boundaries where instead of tightly interlocking, two neighboring faces might be intersecting or slightly apart. If we were to na\"{i}vely to find the closest hits for these degenerate boundaries, this would create incorrect ray segments, which might cause sampling and compositing errors ---casting two front-face culled rays to find an exit and entry point allows us to handle them robustly. 

Although we leverage the hardware acceleration of NVIDIA RTX GPUs in our work, this approach does not explicitly require the usage of OptiX/OWL frameworks or any specialized hardware. One can use any other framework or ray-tracing engine that supports these basic functionalities.


\subsection{Per-Segment Volume Integration}
\label{sec:integrate-segments}


We use linear interpolation to sample elements at equidistant points in a segment. The coefficients to linear interpolation calculations are also utilized to check whether the current element contains the point that needs to be sampled. If not all of these coefficients are between 0 and 1, we keep marching until that becomes the case. When a sample is taken, a transfer function is used to look up its color and transparency, and then it is composited to that segment's color. Marching is terminated when a ray becomes opaque or the next sample position falls behind the exit face.

Our marcher utilizes a method similar to ``Projected Tetrahedra''~\cite{shirley1990rastertet} to determine the exit face for a given element. We do not rasterize the elements directly to the screen, which is more in line with the methods described by Aman et al.~\cite{aman2021bth, aman2021compact} and Sahistan et al.~\cite{sahistan2021Shell}. However, we handle primitives other than tetrahedra as well.

We employ XOR-based compaction schemes to reduce the memory footprint of the data while still allowing efficient traversal. Our compaction process reduces the vertex index storage per element, except hexahedra. We also address memory alignment with this scheme. We store connectivity information in a separate buffer to preserve memory alignment properties. Although reducing memory usage is usually helpful, this index removal may not be desirable for some in~situ scenarios; our marcher can perform without compaction.


Ray-marching processes start from a cluster's shell that contains pbrt-style~\cite{pharr2021pbrtaccel} encoded element information. We can construct the first element behind the shell face using this information. After entering the shell, the connectivity buffer and compacted element information are enough to fetch and construct the next elements along the given ray segment. However, our compaction requires elements to be traversed in sequential order without skipping. When an element is reconstructed from XOR-compacted form, the vertices cannot be in any order because this introduces sampling artifacts. Therefore, our scheme not only re-obtains vertex indices but also consistently places them according to VTK mesh ordering for each element~\cite{VTK}.

An exit face must be determined to select the next element on the ray segment. We project the element vertices to a ray-centric coordinate system to conduct 2D intersection tests to find the exit face. Finally, our method maintains a \emph{march state}, which book keeps the last intersected face type (triangle or quad), current element's type, index, and vertex indices for every marching step. Moreover, we follow a general rule while traversing the volume: placing the entry face indices into the same positions in the march state during marching. This rearrangement of the vertices allows us to ignore the entry face during exit face selection(since we now know which vertices belong to entry face). When the marcher leaps to another element, we update march state old entry face indices with exit face indices. Since each volume element is unique in terms of its geometry and face arrangement, it is hard to make a \emph{simple} algorithm that handles all possible combinations. For this reason, our element marching handles various elements in a case-by-case fashion. 

We handle tetrahedral elements similar to~\cite{sahistan2021Shell}. However, unlike Sahistan et al, we also allow intermediate points inside the elements to be sampled. We transform the vertices to the previously mentioned ray-centric coordinate system to determine the exit face for tetrahedral elements. After each vertex is transformed, we apply a maximum of two 2D left tests to determine the face containing point $(0,0)$.

The exit face can again be found via 2D left tests when inside a pyramid. Due to the quad face that the pyramids have, we utilize the last intersected face type field stored in march state to simplify our left test cases. Using projected vertices, if the entry face is a quad face, we find the exit face among four triangles (similar to the tetrahedron case). Otherwise, we first check if the quad face contains the point $(0, 0)$, then test the remaining three triangles for the same condition. Finally, we update the last intersected face type accordingly.

In terms of finding the exit face, wedges are similar to pyramids. It should be noted that wedges contain three quad faces; hence even if the intersected face type is a quad, the exit face might be another quad face. Like other element types, we ignore the entry face to avoid extra left tests. After the exit face is determined, we update intersected face type once more.

Hexahedra are uniform like tetrahedra; however, they have more faces. Therefore finding the exit intersection requires the highest number of left tests. In the worst case, hexahedra need 13 left tests, whereas wedges, pyramids, and tetrahedra require 7, 5, and 2 left tests, respectively.

\subsection{Deep Compositing}
\label{sec:deep-compositing}

The techniques described in the previous sections allow any rank to efficiently find, for a given ray, all the segments that overlap with that rank's part of the data (\autoref{sec:shell-to-shell}); and to efficiently integrate each of these segments (\autoref{sec:integrate-segments}). This integration step produces one RGB color and opacity value for each segment, plus the depth of the given segment. Borrowing terminology from triangle rasterization, we call each such tuple of color $C$, opacity $\alpha$, and depth $z$, a \emph{fragment} $F=({F_C,\;F_\alpha,\;F_z})$.

\begin{figure}[t]
    \centering
    \includegraphics[width=\columnwidth]{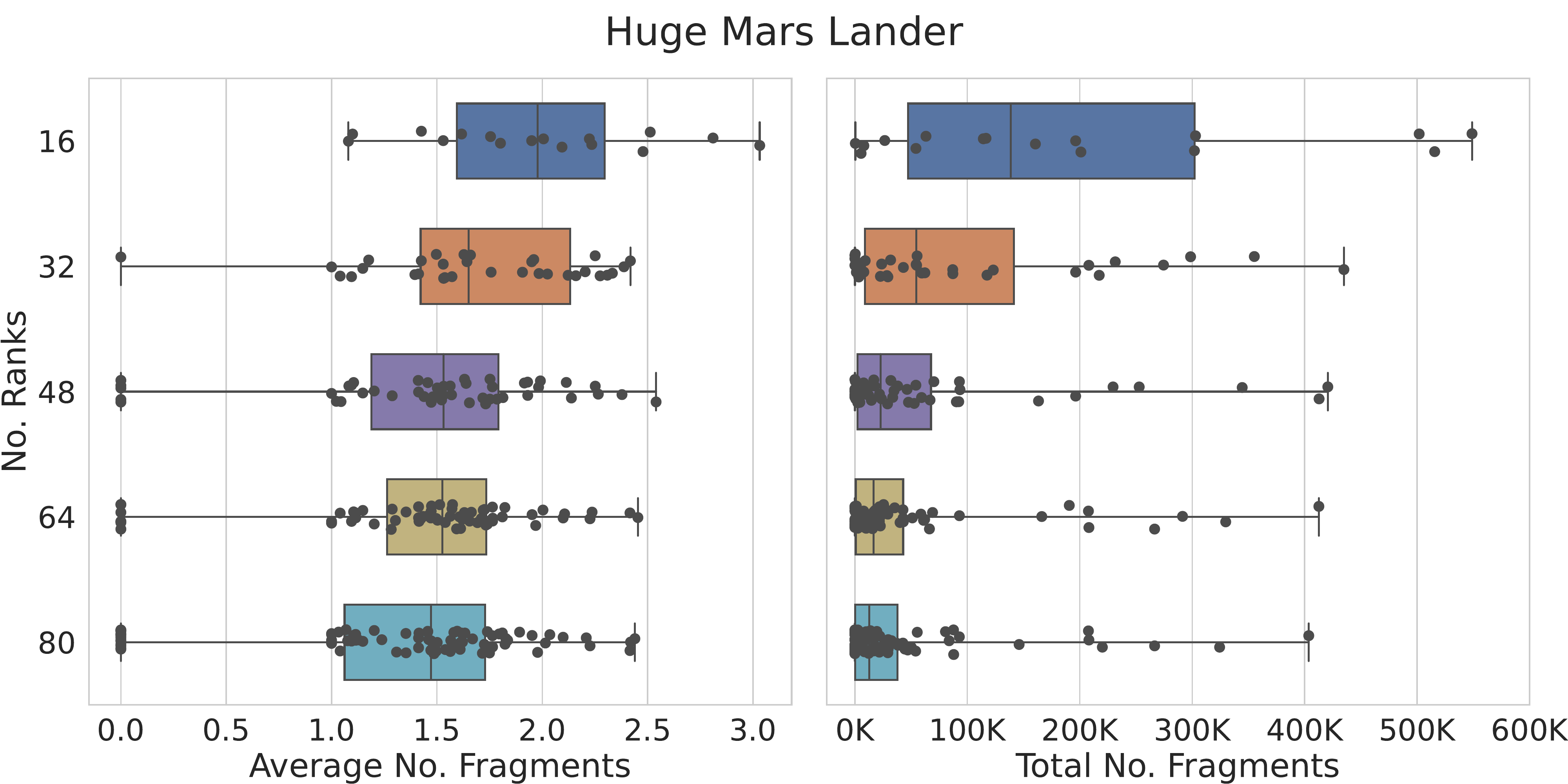}
    \caption{Box plots of average (left) and total (right) number of fragments generated by individual ranks while rendering the Huge Lander. We take averages over non-empty pixels where their opacity is greater than 0. The plots are for the rank counts of 16, 32, 48, 64, and 80. Scattered points signify individual ranks' average (left) and total (right) fragment counts at a given MPI size.}
    \label{fig:num-frags}
\end{figure}

\begin{figure}[tbp]
    \includegraphics[width=.50\columnwidth]{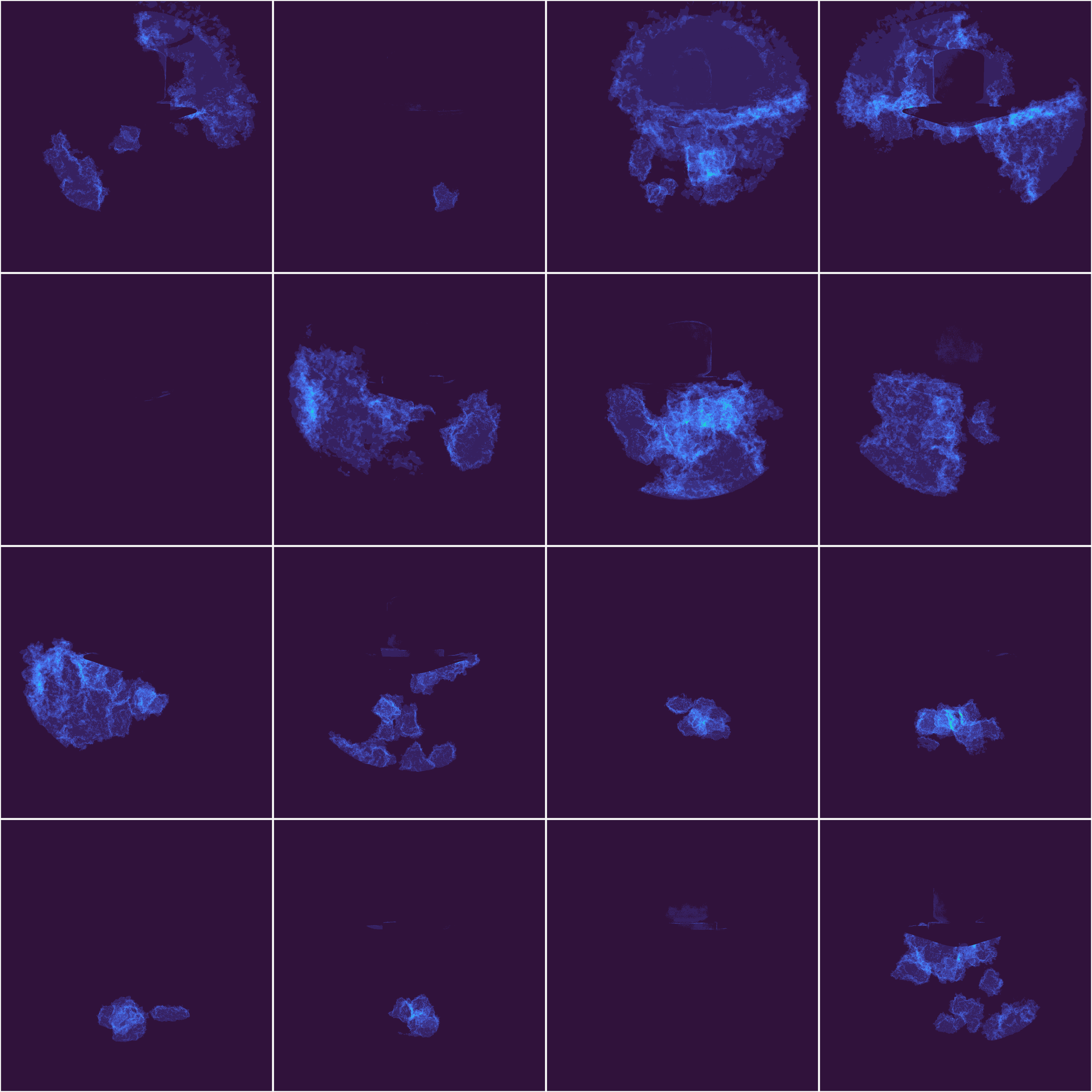} \hspace{-0.38em}
    \includegraphics[width=.50\columnwidth]{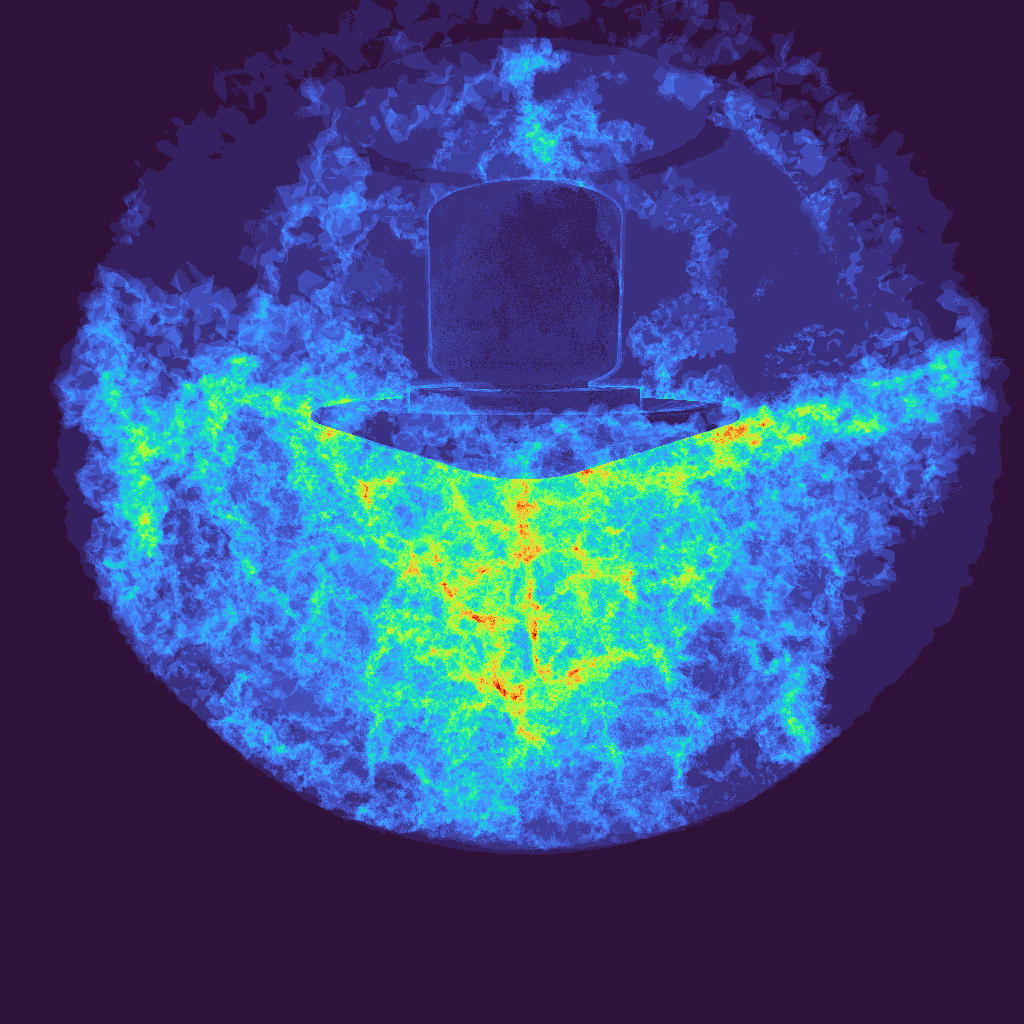}
    \includegraphics[width=1.00\columnwidth]{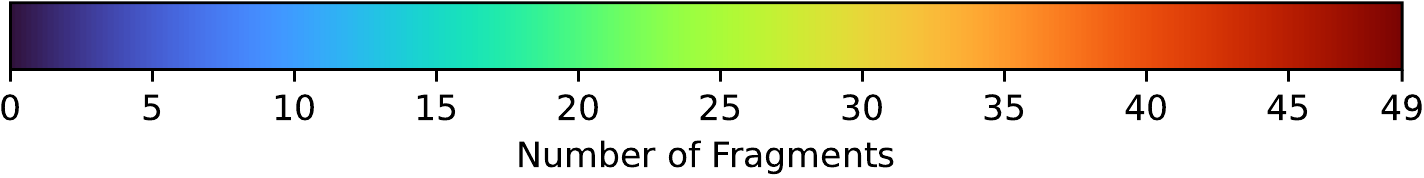}
  \caption{Heatmaps for the number of fragments for a view of the Huge Lander rendered with 16 MPI ranks (first box given in \autoref{fig:num-frags}): On the left, fragments for every 16 rank, and on the right, all heatmaps combined into one image.}
  \label{fig:num-frags-vis}
\end{figure}

Given all of a given pixel $P$'s fragments $F^{(P)}_{0},F^{(P)}_{1},\dots,F^{(P)}_{N^{(P)}}$, the correct final pixel color is the result of first sorting these fragments by their depth and compositing them using
$\widehat{O}(A,B)\;$/$\;\widehat{U}(A,B)$~\cite{OverOperator,Porter84}. The challenge is that any given pixel's fragments may get produced on many different ranks, requiring some merging of different ranks' results. Even worse, the irregular shape of the shells means that any ray can enter and leave the same shell multiple times at multiple distances, producing multiple---and in some cases, many---fragments for the same pixel. \autoref{fig:num-frags} illustrates the distribution of the average and the total number of fragments for a view of the Huge Lander for increasing rank counts. As it can be observed from the case where the rank count is 16 (cf.~\autoref{fig:num-frags-vis}), each pixel may have multiple fragments generated from multiple ranks.

The easiest approach to compositing this would be to first composite all ranks' fragments to a single fragment per pixel per rank and then use some optimized compositing library like IceT~\cite{moreland2011icet} to produce the final image. However, as neither $\widehat{O}$ nor $\widehat{U}$ are commutative, this will give wrong results every time a ray enters the same shell more than once. Fragments need to be composited in visibility order, considering that $N$ fragments along a ray are generally distributed unequally across the ranks. Let $\otimes$ denote compositing operation given any ray that produces two fragments $F^{(A)}_0$ and $F^{(A)}_1$ on the same rank must also have had at least one other fragment $F^{(B)}$ on at least one other rank. This requires compositing as $F^{(A)}_0 \otimes F^{(B)} \otimes F^{(A)}_1$, which in general is different from both $F^{(B)} \otimes (F^{(A)}_0 \otimes F^{(A)}_1)$ and $(F^{(A)}_0 \otimes F^{(A)}_1) \otimes F^{(B)}$.

\subsubsection{Compositing with More than One Fragment/Pixel}

To solve this compositing problem, we developed a new compositing framework that explicitly allows each rank to have multiple fragments per pixel. At an abstract level, our method expects each pixel to store one counter that specifies the number of fragments, $N$, plus an address (or offset) to a list of fragments, $F_0, \; \ldots\;, F_{N-1}$. Similar to parallel-direct-send~\cite{grosset2017todtree, favre2007directsend}, we then split the frame buffer into $R$ distinct \emph{regions} of pixels (where $R$ is the number of ranks); each rank will be responsible for receiving, compositing, and delivering the final composited results of one region of pixels. Compositing then works in the following steps:

\def\compactstep#1{\par\medskip\noindent\textbf{#1}}

\compactstep{1) Generating a contiguous send buffer.} Given each pixel's fragment lists, each rank computes a GPU-parallel prefix sum over all its fragment counts, which also yields the total number of fragments on this rank. We then allocate a single contiguous memory region for these fragments and compact the individual fragments into this buffer (using the prefix sum result as offsets). By design, this buffer will contain all fragments going to all other ranks in order.

\compactstep{2) Exchanging per-pixel fragment counts ranges.} Given the assigned range of pixels, each rank computes which range of per-pixel counters it needs to send to any other rank. To this end, each rank allocates a per-rank counter buffer with size $R$ times the number of pixels in its region. Next, each rank computes the offsets to store the counters from other ranks. We then perform a collective \code{MPI\_Alltoallv} on these buffers, after which each rank has, for its assigned region of pixels, the fragment counts from every other rank.

\compactstep{3) Exchanging Fragment Lists.} Having received all other ranks' per-pixel fragment counts for its range of pixels, each rank then performs a GPU prefix sum over those counters, the result of which can once again be seen as offsets into a compact buffer of all fragments for its range of pixels. Looking up the prefix sums at the correct offsets specifies how many fragments each rank will receive from any other rank and how many fragments it will receive altogether. We then allocate a receiving buffer of the required size, look up where each other rank's fragments will go in this buffer, and issue a second \code{MPI\_Alltoallv} that, in this case, collectively moves all fragments into the receive buffer of the rank assigned to that fragment's corresponding pixels.

\compactstep{4) Local Compositing.} The result of the previous steps is that each rank now has two buffers containing all fragment lists for its assigned pixels. The first buffer ---\textit{fragment buffer}--- stores all fragments for that rank's pixels received from all other ranks, ordered by ranks and pixels within each rank. Given a specific MPI rank, this buffer stores all fragments for that rank's first pixel from rank 0, then all those for its second pixel from rank 0, and so on, followed by all fragments from rank 1, then all fragments from rank 2, and so on. The second buffer ---\textit{offset buffer}--- stores the results of prefix sum operations. It, by design, provides the offsets where the fragment lists start. For example, if $P$ is the number of pixels that this rank is responsible for, then the fragments from rank $r$ for pixel $j$ start at offset \code{offsets[r*P+j]}. Using this, we can now launch a CUDA kernel that, for each pixel $p$, looks up the $R$ different lists of fragments and composites them in the visibility order.

\compactstep{5) Sending final results to master.} The output of the previous CUDA kernel is, on each rank, a fully composited RGBA value for each pixel in that rank's range of pixels. We send these to the master using a \code{MPI\_Send}; the master sets up $R$ matching \code{MPI\_Irecv}s, each of which uses the appropriate part of the final frame buffer as receive buffer. Once these are completed, the master has the final assembled frame buffer, and compositing is complete.

This method is a natural extension of the parallel direct-send technique as described by Grosset et al.~\cite{grosset2017todtree} and Favre et al.~\cite{favre2007directsend}, with the main difference that we not only send one fragment per pixel but variable-sized lists of fragments. We term this method \emph{deep compositing} because it merged the concepts of image-based compositing with the orthogonal concept of \emph{deep frame buffers}~\cite{GershbeinHanrahan00}.

\subsubsection{Fragment List Management}

Though the compositing itself is easy to use from the host side, properly setting up the device-side inputs (fragment lists and counters) would require the renderer to handle what are akin to device-side dynamic memory allocations to manage those per-pixel variable-size fragment lists during rendering.

To relieve the renderer of this low-level fragment list management, we also developed what we call a \emph{device interface} for this library, through which a renderer can simply \emph{write} new fragments into a pixel, with that interface then handling the proper storage of those fragments---which significantly simplifies the rendering code.

\paragraph{Two-Pass, Flexible-length Fragment Lists}

The main challenge for developing this interface was that we could not simply allocate more device memory during rendering, so we needed \emph{some} limit on how many fragments a renderer would be allowed to generate in any frame. We first developed a two-stage interface in which the renderer would be run twice: in the first stage, the interface would only count the fragments produced per pixel but not store any. After this stage, it would compute a prefix sum over those counters to allocate a big enough buffer, with the prefix sum values serving as offsets into this buffer. A second pass would then perform exactly the same rendering but store the fragments at the provided offsets.

\paragraph{Single-Pass, Fixed-Length Fragment Lists}

The two-pass method allows for arbitrary-sized fragment lists (up to device memory, obviously); but requires running at least the shell traversal twice, which may or may not be acceptable. We, therefore, also developed a second, single-pass device interface in which the renderer---upon initialization---specifies a maximum allowed number of fragments per pixel, which can then be used to pre-allocate lists to add fragments. Having a single pass is straightforward but requires some form of \emph{overflow}-handling if a render wants to submit fragments to a pixel whose list is already full. We currently implement two methods for this overflow handling: In the \emph{drop} method, we perform insertion sort into the existing list and simply drop the latest fragment. In \emph{merge}, we find the fragment with the lowest opacity and perform a \emph{over} compositing of this element onto the one in front of it (i.e., using the depth from the previous one), then insert the new fragment into the list.

\subsubsection{Implementation Details}

Though primarily developed for this particular application, we believe the method just described is also applicable to other, similar applications, and thus decided to implement this into a stand-alone \emph{deep compositing} library that the rest of our renderer then uses. The compositing itself uses MPI, for which in our application, we use a CUDA-aware version of OpenMPI 4.1. Using CUDA-aware MPI allows the compositing code to directly operate on device buffers, which means that the same library can work with both host and device-side renderers. We use CUDA for the device interface, with a simply host-side interface to initialize and trigger compositing.

The bandwidth required for compositing is often a bottleneck in data-parallel parallel rendering, even with only a single fragment per pixel. One step we use to reduce bandwidth is that we allow the user to specify whether to use full \emph{float} precision for fragments (five floats total, for r, g, b, depth, and opacity); or to use a lower-precision encoding with 8-bit fixed-point for RGBA, and floats only for the depth value. The device interface in both cases is the same, but that interface encodes fragments as they get submitted. We also automatically discard fragments with zero opacity value, as these will not contribute to the image.

Aside from the fragments, the per-pixel counters require a large bandwidth. To reduce that, we use specialized encodings with 2, 4, 8, or 32 bits for those counters, depending on the longest fragment list length. We use dedicated CUDA kernels for encoding and decoding the 32-bit counter arrays into this lower-precision representation before and after the \code{MPI\_Alltoallv} counter exchange; otherwise, perform the algorithm exactly as described above.

\section{Experimental Results}
\label{sec:evaluation}

We conducted our experiments on Frontera RTX nodes of Texas Advanced Computing Center (TACC), where each of the 22 nodes had four NVIDIA Quadro RTX 5000 plugged into it. We utilize all four GPUs available per node for every data point of our experiments. 

\subsection{Evaluation of the Framework}

We evaluate our rendering framework on Small Mars Lander and Huge Mars Lander data sets. \autoref{fig:teaser} show images of the Small Mars Lander rendered using our framework. Since Small Mars Lander has 72 clusters, we evaluate our framework using 72 GPUs distributed over 18 compute nodes where each cluster is loaded on a separate GPU. For Small Mars Lander, we achieve our peak performance using 72 GPUs yielding the average fps of 14.35. Since the TACC supercomputer does not have more than 22 RTX nodes, we could not test one cluster per GPU scenario for the Huge Mars Lander data set. Therefore, we scale up to a maximum GPU count of 88, yielding 9.83 fps. However, we observe our average peak performance of 10.25 fps for the Huge Mars Lander at 80 GPUs. 

\begin{figure}[htbp]
    \begin{center}
    \includegraphics[width=0.44\columnwidth]{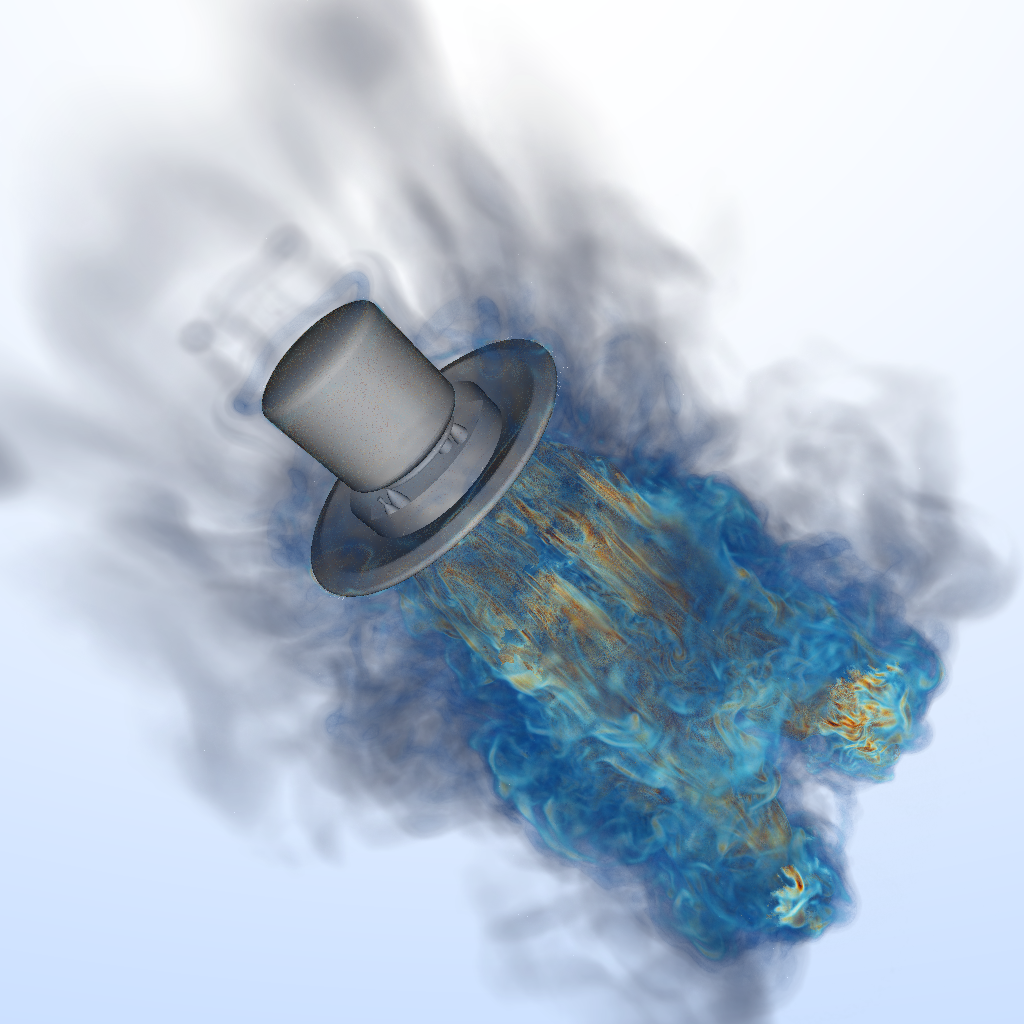}
    \includegraphics[width=0.44\columnwidth]{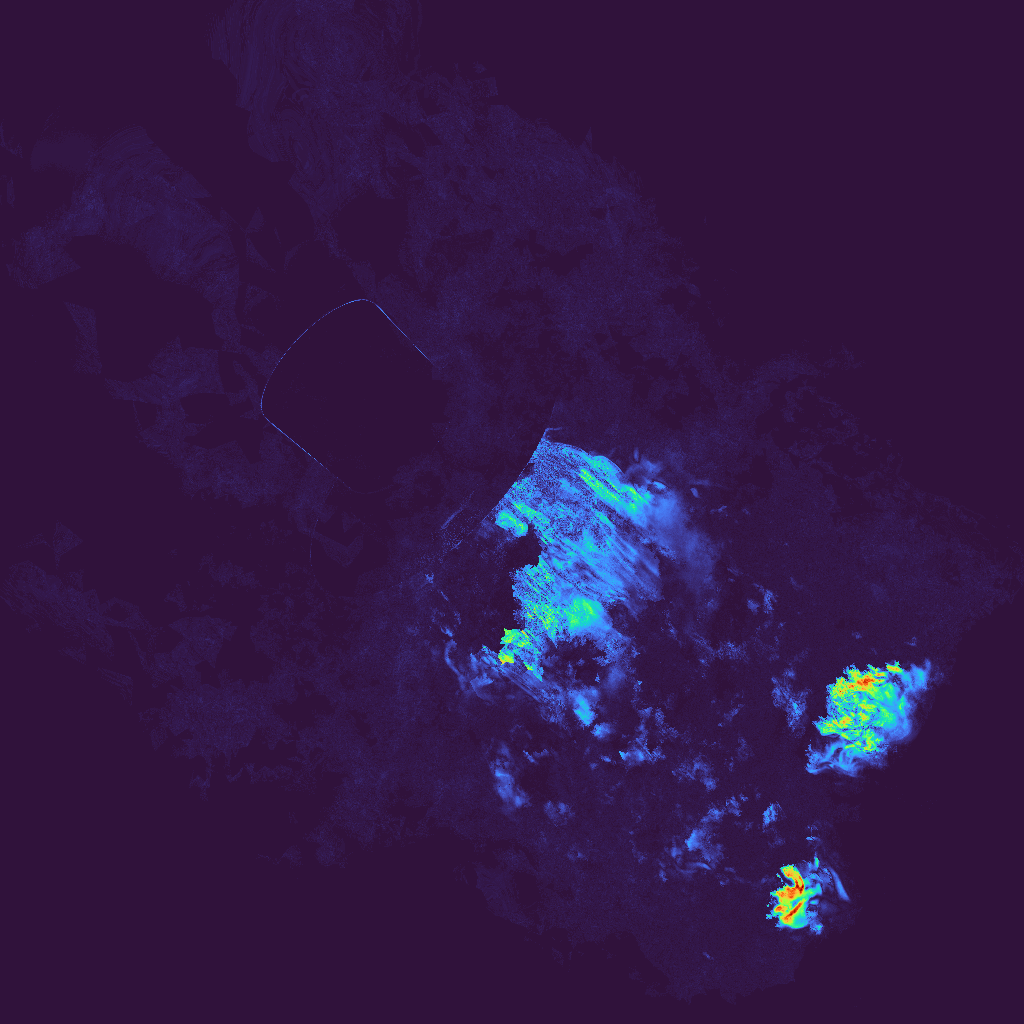}
    \includegraphics[height=0.44\columnwidth]{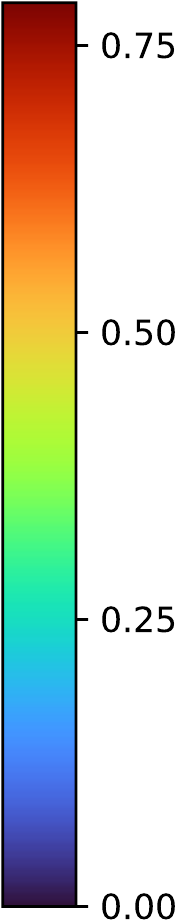}
    \end{center}
    \caption{The left image is the rendering with single fragment compositing (similar to IceT \cite{moreland2011icet}). The right image is the heatmap showing the L2 difference between single fragment compositing and our deep compositing.}
    \label{fig:ice_t_diff}
\end{figure}

Moreover, we evaluate our deep compositing scheme's correctness compared to a single fragment compositing technique. \autoref{fig:ice_t_diff} shows an image rendered by single fragment compositing and a heatmap that compares the difference between single fragment compositing and our deep compositing. The single fragment compositing method depicted is similar to the image-based single image per node compositing techniques such as IceT~\cite{moreland2011icet}.

\subsection{Memory Overhead}

\begin{figure*}[htbp]
    \begin{center}
    \fbox{%
    \begin{minipage}{.48\textwidth}
    \includegraphics[align=c, width=.40\columnwidth]{png/offlineViewer_frame18.png}
    \hfill
    \includegraphics[align=c, width=.58\columnwidth]{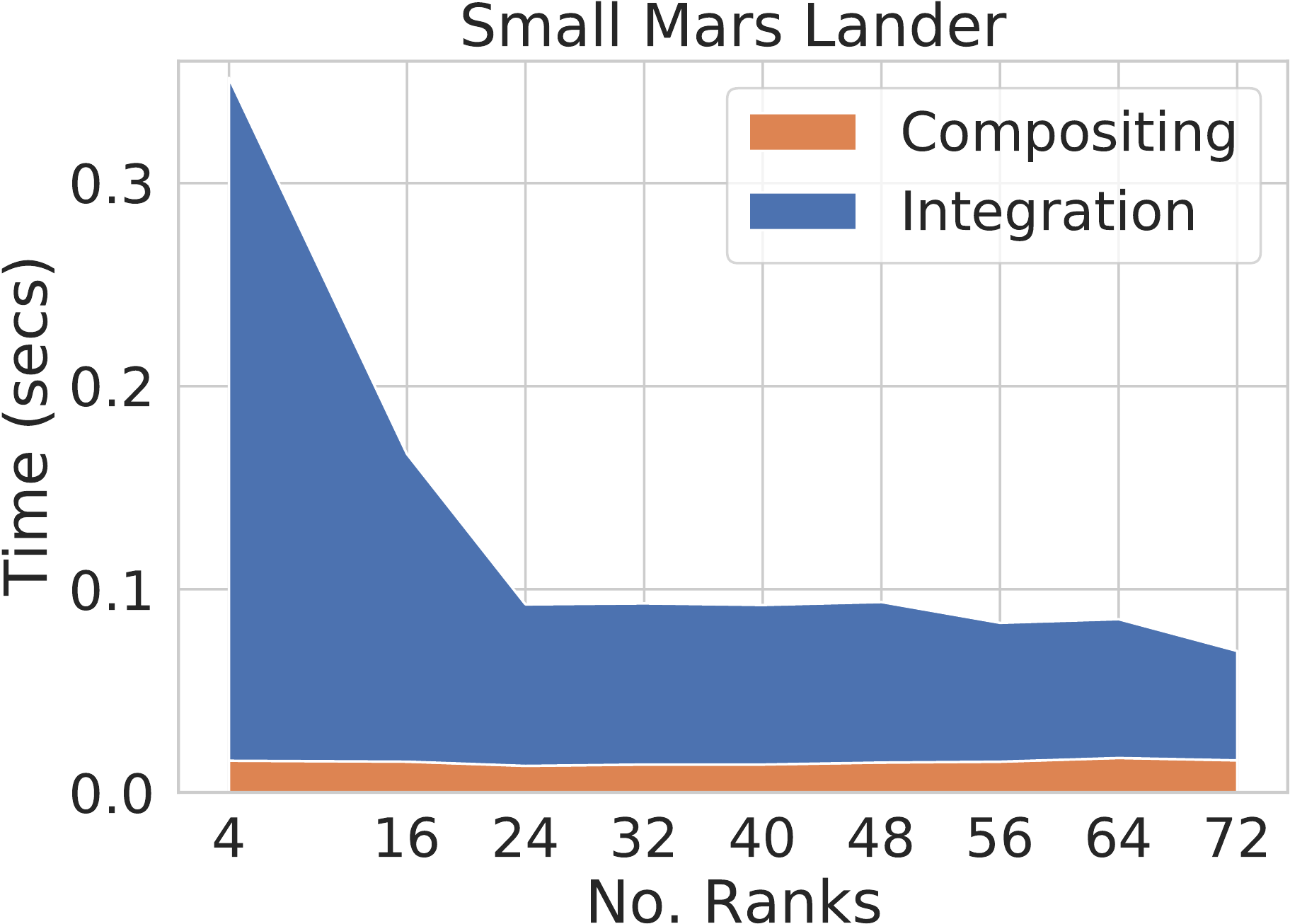}
    \end{minipage}
    }%
    \hfill
    \fbox{%
    \begin{minipage}{.48\textwidth}
    \includegraphics[align=c, width=.40\columnwidth]{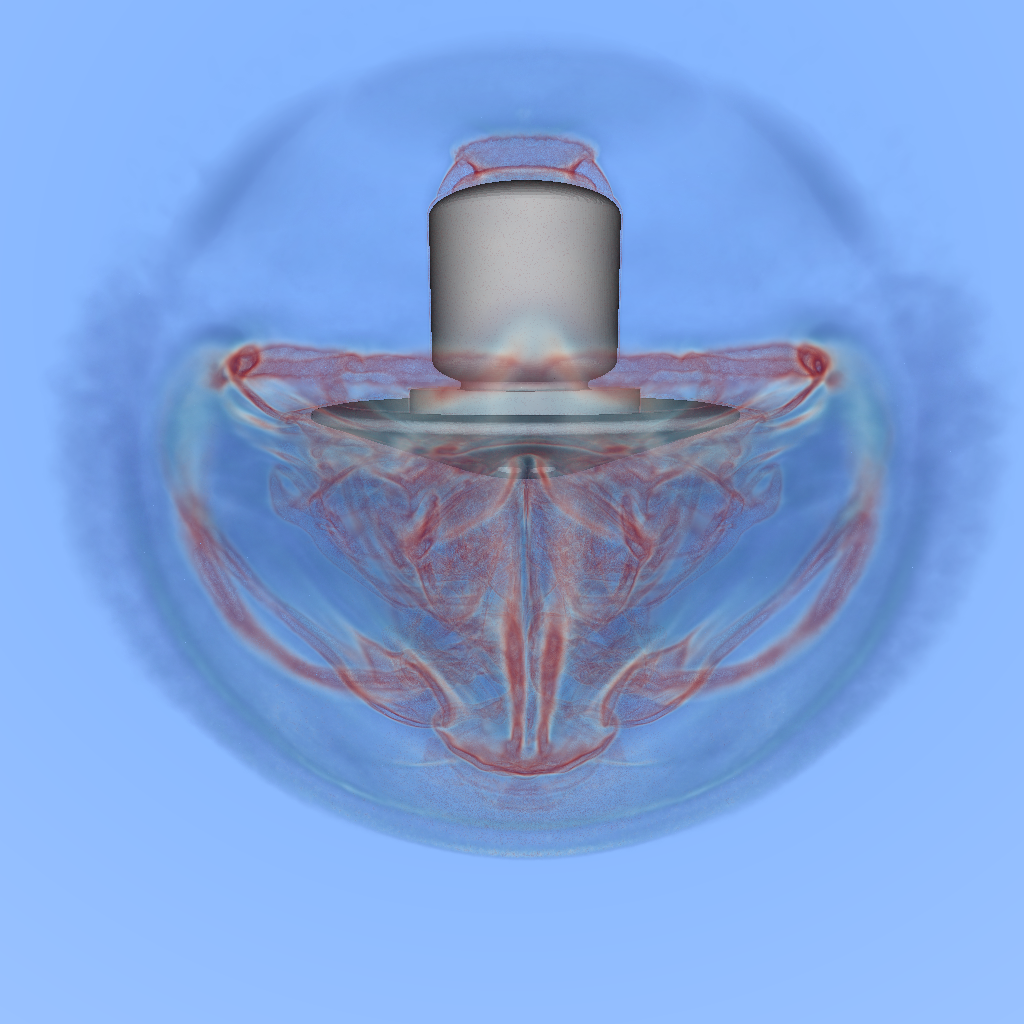}
    \hfill
    \includegraphics[align=c, width=.58\columnwidth]{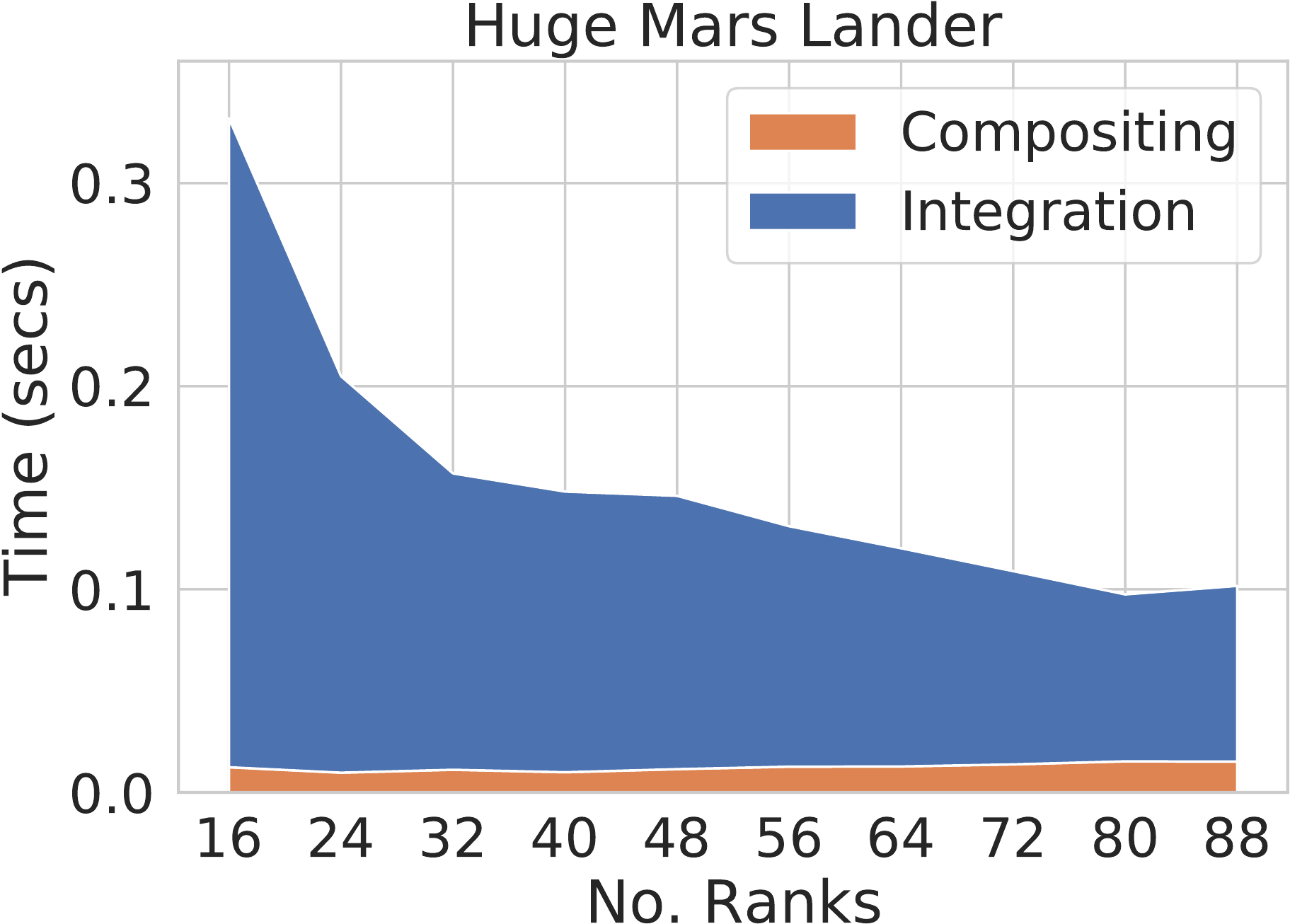}
    \end{minipage}
    }%
    \end{center}
    \caption{Results of the scalability benchmarks for Small Mars Lander (left) and Huge Mars Lander (right) on RTX nodes of the Frontera system on TACC. Integration process timings are stacked over compositing process timings for the given number of ranks to form total rendering times.
    }
    \label{fig:timing_plots}
\end{figure*}

\begin{figure}[htbp]
    \begin{center}
    \includegraphics[width=.495\columnwidth]{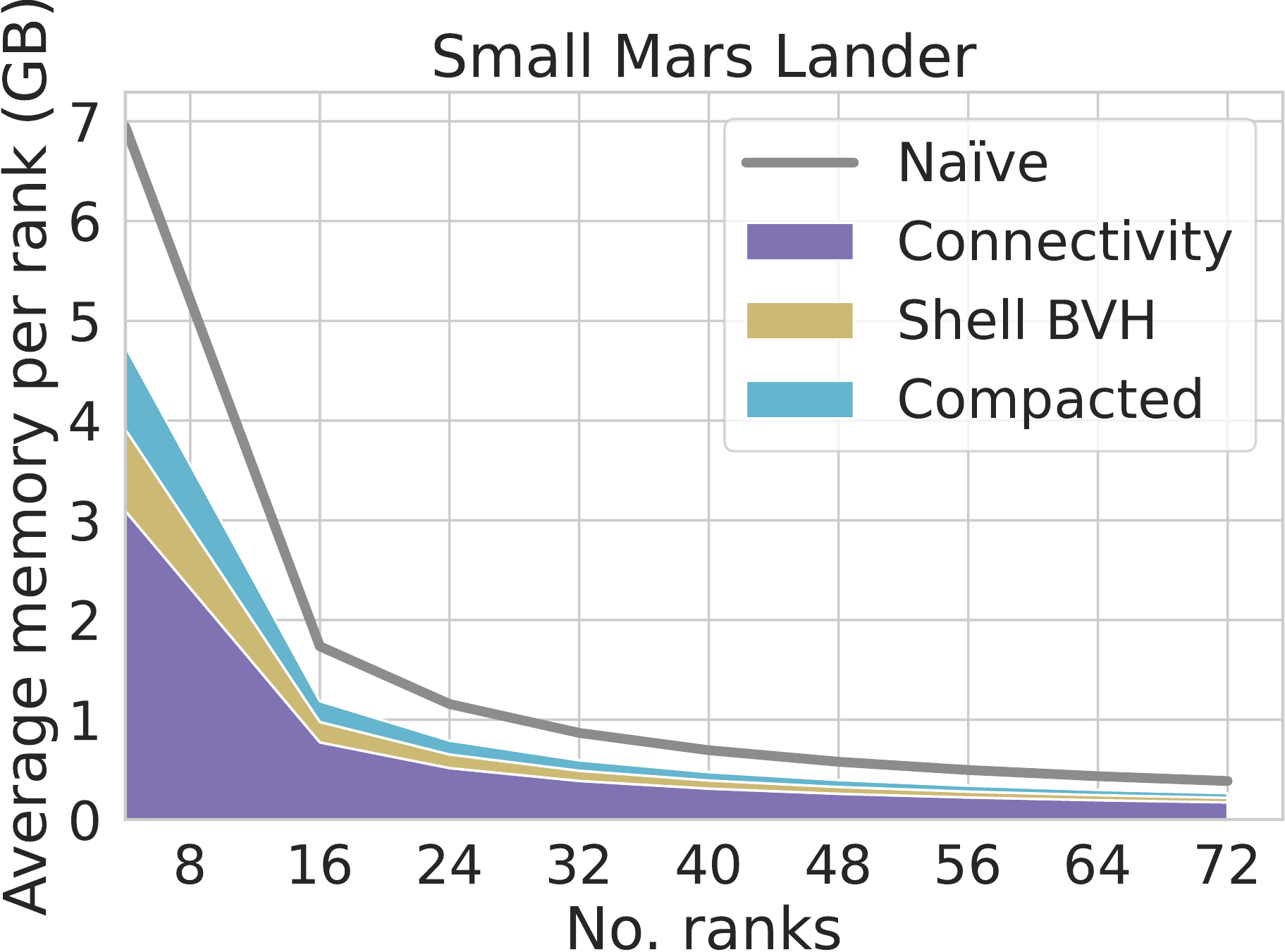}
    \includegraphics[width=.495\columnwidth]{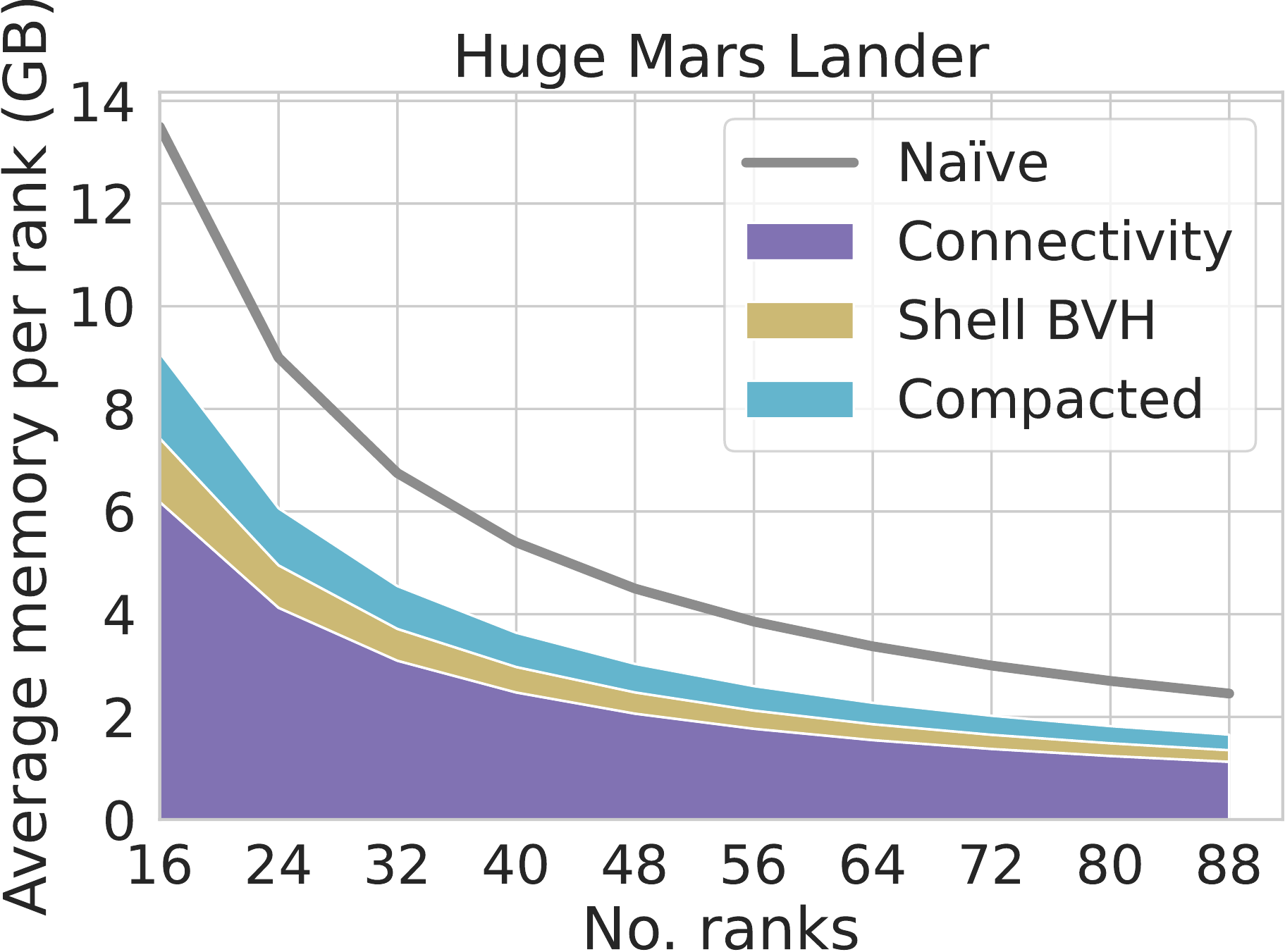}
    \end{center}
    \caption{Per rank average memory consumption of various buffers for increasing MPI sizes: Small Mars Lander (left) and Huge Mars Lander (right). Average memory usage of XOR-compacted elements is stacked over average shell-BVH size, which again is stacked over average connectivity buffer size, providing the total memory usage introduced by these data. We also include a line that indicates the per rank average memory usage without XOR-based compaction (size of Shell-BVH + connectivity buffer + non-compact elements).}
    \label{fig:memory_plots}
\end{figure}

We examine data distribution and memory footprints. \autoref{tab:element_shell_stats} displays the minimum, maximum and average counts per rank of volume elements and shell faces for our two data sets. The MPI sizes (rank counts) in the table are the sizes that experience the highest level of changes in terms of rendering times presented in \autoref{fig:timing_plots}. \autoref{fig:memory_plots} illustrates average memory footprints of our large data structures that may not be present in a simulation environment. Although connectivity information will likely be in most of the simulation systems, we wanted to include connectivity here for the sake of simulation systems like~\cite{ishii20194dtreebased}. 

\begin{table}[htbp]
\centering
\caption{Per rank statistics for selected MPI sizes for two test data sets. The first column gives the MPI size with a number and data set as ``small'' for Small Mars Lander and ``huge'' for Huge Mars Lander. Columns depict minimum, maximum, and average volume element counts and minimum, maximum, and average shell face counts per rank.}
{\small
\begin{tabular}{cccc|ccc}
\hline
\multicolumn{1}{c}{\multirow{2}{*}{\begin{tabular}[c]{@{}c@{}}MPI Size\\ \& data set\end{tabular}}} &
  \multicolumn{3}{c|}{Elements} &
  \multicolumn{3}{c}{Shell} \\ \cline{2-7} 
\multicolumn{1}{c}{} &
  \multicolumn{1}{c}{min.} &
  \multicolumn{1}{c}{max.} &
  \multicolumn{1}{c|}{avg.} &
  \multicolumn{1}{c}{min.} &
  \multicolumn{1}{c}{max.} &
  \multicolumn{1}{c}{avg.} \\
  \hline 
 4-small  & 192.2 M & 203.4 M & 199.6 M & 14.4 M & 16.2 M & 15.2 M \\
24-small & 30.8 M  & 34.9 M  & 33.3 M  & 2.0 M  & 2.9 M  & 2.5 M  \\
40-small & 10.0 M  & 23.3 M  & 20.0 M  & 0.8 M  & 1.9 M  & 1.5 M  \\
72-small & 9.9 M & 11.9 M  & 11.1 M  & 0.6 M  & 1.1 M  & 0.8 M  \\ \hline
16-huge & 365.2 M & 414.6 M & 399.2 M & 21.4 M & 25.1 M & 23.0 M \\
32-huge & 177.8 M & 213.8 M & 200.0 M & 10.3 M & 13.2 M & 11.5 M \\
56-huge & 93.8 M & 121.0 M & 114.1 M & 5.4 M  & 7.3 M  & 6.6 M  \\
88-huge & 60.9 M  & 85.1 M  & 72.6 M  & 3.3 M  & 5.3 M  & 4.2 M \\ \hline
\end{tabular}
}
\label{tab:element_shell_stats}
\end{table}

\subsection{Scalability}

To assess the scalability, we test our approach for increasing the number of ranks (MPI sizes). We also measure sub-process timings, specifically the segment integration and compositing times. We compute them by using \code{MPI\_Barrier}s before and after integration calls to synchronize the processes before starting and ending the timers. Since these barriers stall early terminating integration processes, it causes the total rendering time to increase. For this reason, we measure the total rendering time and integration time in different runs and derive the compositing time by subtracting the total time from the integration time. We calculate the timings reported in \autoref{fig:timing_plots} as follows: we take an average for 20 sequential timesteps of the selected scalar field over 30 runs. Then we take the mean of these 20 average values to form the data points for the given MPI sizes.

\subsection{Fragment Distribution}

Since it is crucial to assess workload distribution across compute nodes for both compositing and integration steps, we measure fragment counts generated by each GPU during ray segment generation. To this end, we present a box plot in \autoref{fig:num-frags}, which depicts the total and average fragment counts for specific MPI sizes on the Huge Mars Lander. We also visualize a series of ``heat images'' for the case where the rank count is equal to 16, which illustrates the fragments composited for a view of the Huge Mars Lander. Each small image shows given nodes generated fragment counts, and the final image accumulates them on top of each other.

\section{Discussion}
\label{sec:discussion}

We evaluate our approach regarding memory consumption, rendering correctness, and scalability. Among the data we precompute and store, the connectivity information takes up the lion's share with ratios around $\approx62.72\%$ for our test cases. We expect this since our approach stores one integer for all faces of a given volume element. XOR-compacted volume elements are the second-largest structure with ratios around $\approx17.11\%$, closely followed by Shell-BVH sizes ($\approx12.74\%$) of the total memory consumption. We observe that the total memory footprint of XOR-compacted representations is $\approx72.49\%$ smaller than their uncompacted versions. The presented framework is memory-wise compatible with in~situ scenarios because many modern simulation systems already store connectivity information. Our XOR-compaction reduces the space required for geometry information, and our shell-BVH sizes stay relatively small despite the large counts of shell faces.







We observe that simple image compositing is not an option for non-trivially partitioned data sets like we present. As the error metric in \autoref{fig:ice_t_diff} confirms, single image compositing gives inaccurate results. Moreover, we found our compositing process to create low overheads even with the GPU counts going up to 88. Although the communication cost for compositing created an increasing trend in terms of time, it never surpassed 22.68\% of the total rendering time, which is an indication that our deep compositing method is highly scalable and generates correctly composited images interactively (see~\autoref{fig:timing_plots}). 

Although the proposed ray-marcher is suitable for the use-cases described, any other ray-marcher that can adequately handle non-convex boundaries can be utilized. For instance, point-query sampling techniques that can leverage adaptive sampling or space skipping~\cite{morrical2019spaceskip,Wang2020noveladaptivesampling} may produce much faster results. However, such methods rely heavily on hierarchical data structures to sample the volume, which would create more additional memory overhead. One could claim point-query sampling techniques negate this memory over-head by not storing connectivity; however, as pointed out before, many simulation environments have that data out of the box. Furthermore, point-query sampling techniques usually produce a noisy image that requires some convergence time to be passed, whereas our marching method generates deterministic noise-free images. For these reasons, we consider our marching algorithm to be more pragmatic in the context of data-parallel rendering and deep compositing.

Finally, examining data distributions from \autoref{tab:element_shell_stats}, we see that directly utilizing native partitioning of the data causes uneven load balancing for some MPI sizes. \autoref{fig:num-frags} reveals this phenomenon where the average number of fragments per rank distribution varies. The effects of this phenomenon can also be observed in \autoref{fig:num-frags-vis}, where ranks 1, 4, and 14 have significantly fewer fragments than the others. Even though native partitioning causes uneven workloads, our timing experiments (cf.~\autoref{fig:timing_plots}) display decent scalability with the increasing number of GPUs we utilized. We smoothly achieve interactive rates with both of our data sets. For the small Mars Lander that has 72 clusters, we benchmark \emph{14.35 fps} using 72 GPUs, and for the 552-cluster huge Mars Lander, we measure \emph{10.27 fps} using 80 GPUs. We also observe an ongoing downwards trend for the timings with the increasing number of GPUs, so it is worth mentioning that our application can achieve even higher frame rates given a more extensive hardware setup. 

Furthermore, it is clear from \autoref{fig:timing_plots} that dominating term of rendering times is volume integration via ray-marching. Nevertheless, we observe a sharp increase in integration performance at $n=24$ and $n=32$ for Small and Huge Mars Lander, respectively. At the same time, it is expected for an embarrassingly parallel ray-casting algorithm to get faster with the increasing number of ranks; it is also likely for a compositing algorithm to slow down due to communication costs. We observe little to no increase in timings with our deep compositing, where it nearly behaves like a constant.






\section{Conclusions and Future Work}
\label{sec:conclusion}

We introduce a GPU-based direct volume visualization framework that allows correct and interactive rendering even for non-convexly partitioned data. Our framework presents a mixed element ray-marching algorithm to integrate ray segments along the viewing direction. We achieved memory savings by exploiting XOR-based compaction schemes on our finite element data structures. Furthermore, we illustrate a deep compositing algorithm that allows proper order compositing of the RGBA-Z values obtained across multiple compute nodes. Our framework scales well for increasing GPU counts while using native partitioning of non-convex data sets. We consider our framework suitable for both in~situ and post~hoc applications.


Possible areas for further research are as follows. While we allow visualizations of multiple scalar fields and timesteps, we do not use double/triple buffering techniques that can hide the buffer loading times. Our implementation naively takes one scalar set on request (i.e., does not pre-fetch anything). In order to improve the time steps loading performance, buffering and pre-fetching the time steps in GPU and main memory can be employed \cite{shih2014out}. Moreover, currently, we assume that the topology of the volumetric data does not change through time, yet this may not be the case. Furthermore, our sampling method does not support bilinear elements since determining vertex index order after element construction is difficult for them using our XOR-compaction. Also, we would utilize another compaction scheme over connectivity information as other works do \cite{muigg2011interactivevol, aman2021compact}. Image-based partitioning may further increase our method's efficiency; however, it can get challenging with the in~situ emphasis. Finally, integrating our approach into existing frameworks is another future work, field testing our claims.

\acknowledgments{
Intentionally omitted for review.}

\balance
\bibliographystyle{abbrv-doi}
\bibliography{lander}

\begin{thebibliography}{10}

\bibitem{abram2018galaxy}
G.~Abram, P.~Navrátil, P.~Grossett, D.~Rogers, and J.~Ahrens.
\newblock Galaxy: Asynchronous ray tracing for large high-fidelity
  visualization.
\newblock In {\em Proceedings of the IEEE 8th Symposium on Large Data Analysis
  and Visualization}, LDAV~'18, pp. 72--76, 2018.

\bibitem{aman2021compact}
A.~Aman, S.~Demirci, and U.~G{\"u}d{\"u}kbay.
\newblock Compact tetrahedralization-based acceleration structures for ray
  tracing.
\newblock {\em Journal of Visualization}, In press.

\bibitem{aman2021bth}
A.~Aman, S.~Demirci, U.~G\"{u}d\"{u}kbay, and I.~Wald.
\newblock Multi-level tetrahedralization-based accelerator for ray-tracing
  animated scenes.
\newblock {\em Computer Animation and Virtual Worlds}, 32(3-4):e2024, 11 pages,
  2021.

\bibitem{Anderson99}
W.~K. Anderson, W.~D. Gropp, D.~K. Kaushik, D.~E. Keyes, and B.~F. Smith.
\newblock Achieving high sustained performance in an unstructured mesh {CFD}
  application.
\newblock In {\em Proceedings of the ACM/IEEE Conference on Supercomputing}, SC
  '99, p. 69–es. Association for Computing Machinery, New York, NY, USA,
  1999.

\bibitem{aupy2019highthroughputinsitu}
G.~Aupy, B.~Goglin, V.~Honoré, and B.~Raffin.
\newblock Modeling high-throughput applications for in situ analytics.
\newblock {\em The International Journal of High Performance Computing
  Applications}, 33(6):1185--1200, 2019.

\bibitem{ayachit2015catalyst}
U.~Ayachit, A.~Bauer, B.~Geveci, P.~O'Leary, K.~Moreland, N.~Fabian, and
  J.~Mauldin.
\newblock Paraview catalyst: Enabling in situ data analysis and visualization.
\newblock In {\em Proceedings of the First Workshop on In Situ Infrastructures
  for Enabling Extreme-Scale Analysis and Visualization}, ISAV2015, p. 25–29.
  Association for Computing Machinery, New York, NY, USA, 2015.

\bibitem{biedert2018hwacceleratedmultitile}
T.~Biedert, P.~Messmer, T.~Fogal, and C.~Garth.
\newblock Hardware-accelerated multi-tile streaming for realtime remote
  visualization.
\newblock In H.~Childs and F.~Cucchietti, eds., {\em Proceedings of the
  Eurographics Symposium on Parallel Graphics and Visualization}. The
  Eurographics Association, 2018.

\bibitem{biedert2017taskbasedparallel}
T.~Biedert, K.~Werner, B.~Hentschel, and C.~Garth.
\newblock A task-based parallel rendering component for large-scale
  visualization applications.
\newblock In A.~Telea and J.~Bennett, eds., {\em Eurographics Symposium on
  Parallel Graphics and Visualization}. The Eurographics Association, 2017.

\bibitem{binyabib2019hybrid}
R.~Binyahib, T.~Peterka, M.~Larsen, K.-L. Ma, and H.~Childs.
\newblock A scalable hybrid scheme for ray-casting of unstructured volume data.
\newblock {\em IEEE Transactions on Visualization and Computer Graphics},
  25(7):2349--2361, 2019.

\bibitem{brownlee2013imageparallel}
C.~Brownlee, T.~Ize, and C.~D. Hansen.
\newblock Image-parallel ray tracing using {OpenGL} interception.
\newblock In F.~Marton and K.~Moreland, eds., {\em Eurographics Symposium on
  Parallel Graphics and Visualization}. The Eurographics Association, 2013.

\bibitem{cao2019parallelvis}
Y.~Cao, Z.~Mo, Z.~Ai, H.~Wang, and Z.~Zhang.
\newblock Parallel visualization of large-scale multifield scientific data.
\newblock {\em Journal of Visualization}, 22, 08 2019.

\bibitem{castanie2006distributedsharedmemory}
L.~Castanie, C.~Mion, X.~Cavin, and B.~Levy.
\newblock Distributed shared memory for roaming large volumes.
\newblock {\em IEEE Transactions on Visualization and Computer Graphics},
  12(5):1299--1306, 2006.

\bibitem{childs2006hybridmassive}
H.~Childs, M.~A. Duchaineau, and K.~Ma.
\newblock A scalable, hybrid scheme for volume rendering massive data sets.
\newblock In A.~Heirich, B.~Raffin, and L.~P.~P. dos Santos, eds., {\em
  Proceedings of the 6th Eurographics Symposium on Parallel Graphics and
  Visualization, EGPGV@EuroVis/EGVE 2006, Braga, Portugal, May 11-12, 2006},
  pp. 153--161. Eurographics Association, 2006.

\bibitem{childs2020terminsitu}
H.~Childs~et al.
\newblock A terminology for in situ visualization and analysis systems.
\newblock {\em The International Journal of High Performance Computing
  Applications}, 34(6):676--691, 2020.

\bibitem{demarle2021temporalinsitu}
D.~E. DeMarle and A.~C. Bauer.
\newblock \textit{In Situ} visualization with temporal caching.
\newblock {\em Computing in Science Engineering}, 23(3):25--33, 2021.

\bibitem{favre2007directsend}
S.~Eilemann and R.~Pajarola.
\newblock Direct send compositing for parallel sort-last rendering.
\newblock In J.~M. Favre, L.~P. Santos, and D.~Reiners, eds., {\em Proceedings
  of the Eurographics Symposium on Parallel Graphics and Visualization},
  EGPGV~07. The Eurographics Association, 2007.

\bibitem{garrity1990rtirregvolume}
M.~P. Garrity.
\newblock Raytracing irregular volume data.
\newblock {\em ACM Computer Graphics (Proceedings of SIGGRAPH~90)},
  24(5):35–40, 1990.

\bibitem{GershbeinHanrahan00}
R.~Gershbein and P.~Hanrahan.
\newblock A fast relighting engine for interactive cinematic lighting design.
\newblock In {\em Proceedings of the 27th Annual Conference on Computer
  Graphics and Interactive Techniques}, SIGGRAPH '00, p. 353–358. ACM
  Press/Addison-Wesley Publishing Co., USA, 2000.

\bibitem{grosset2016imagecompositing}
A.~V.~P. Grosset, A.~Knoll, and C.~Hansen.
\newblock Dynamically scheduled region-based image compositing.
\newblock In {\em Proceedings of the Eurographics Symposium on Parallel
  Graphics and Visualization}, vol. 2016 of {\em EGPGV~'16}, 6 2016.

\bibitem{grosset2017todtree}
A.~V.~P. Grosset, M.~Prasad, C.~Christensen, A.~Knoll, and C.~Hansen.
\newblock {TOD-Tree: Task-Overlapped Direct} send {Tree} image compositing for
  hybrid {MPI} parallelism and {GPUs}.
\newblock {\em IEEE Transactions on Visualization and Computer Graphics},
  23(6):1677--1690, 2017.

\bibitem{hadwiger2018sparseleap}
M.~Hadwiger, A.~K. Al-Awami, J.~Beyer, M.~Agus, and H.~Pfister.
\newblock \textit{SparseLeap}: Efficient empty space skipping for large-scale
  volume rendering.
\newblock {\em IEEE Transactions on Visualization and Computer Graphics},
  24(1):974--983, 2018.

\bibitem{OverOperator}
M.~Ikits, J.~Kniss, A.~Lefohn, and C.~Hansen.
\newblock Volume rendering techniques.
\newblock In R.~Fernando, ed., {\em {GPU} Gems}, pp. 667--692. Addison-Wesley,
  2004.
\newblock Available at
  \url{https://developer.nvidia.com/sites/all/modules/custom/gpugems/books/GPUGems/gpugems_ch39.html},
  Accessed: 27 March 2022.

\bibitem{ishii20194dtreebased}
M.~Ishii, M.~Fernando, K.~Saurabh, B.~Khara, B.~Ganapathysubramanian, and
  H.~Sundar.
\newblock Solving {PDEs} in space-time: {4D} tree-based adaptivity, mesh-free
  and matrix-free approaches.
\newblock In {\em Proceedings of the International Conference for High
  Performance Computing, Networking, Storage and Analysis}, SC '19, Article no.
  61, 22 pages. Association for Computing Machinery, New York, NY, USA, 2019.

\bibitem{kruger2003accelerationgpu}
J.~Kruger and R.~Westermann.
\newblock Acceleration techniques for {GPU-based} volume rendering.
\newblock In {\em Proceedings of IEEE Visualization}, VIS~03, pp. 287--292,
  2003.

\bibitem{larsen2017alpine}
M.~Larsen, J.~Ahrens, U.~Ayachit, E.~Brugger, H.~Childs, B.~Geveci, and
  C.~Harrison.
\newblock {The ALPINE} in situ infrastructure: {Ascending from the ashes of
  Strawman}.
\newblock In {\em Proceedings of the In Situ Infrastructures on Enabling
  Extreme-Scale Analysis and Visualization}, ISAV'17, p. 42–46. Association
  for Computing Machinery, New York, NY, USA, 2017.

\bibitem{Larsen2015strawman}
M.~Larsen, E.~Brugger, H.~Childs, J.~Eliot, K.~Griffin, and C.~Harrison.
\newblock Strawman: A batch in situ visualization and analysis infrastructure
  for multi-physics simulation codes.
\newblock In {\em Proceedings of the First Workshop on In Situ Infrastructures
  for Enabling Extreme-Scale Analysis and Visualization (ISAV), held in
  conjunction with SC15}, pp. 30--35. Austin, TX, Nov. 2015.

\bibitem{larsen2015raytracingdataparallel}
M.~Larsen, J.~S. Meredith, P.~A. Navrátil, and H.~Childs.
\newblock Ray tracing within a data parallel framework.
\newblock In {\em Proceedings of IEEE Pacific Visualization Symposium},
  PacificVis~15, pp. 279--286, 2015.

\bibitem{liukwan1995parallelvolumedistributed}
K.-L. Ma.
\newblock Parallel volume ray-casting for unstructured-grid data on
  distributed-memory architectures.
\newblock In {\em Proceedings of the IEEE Symposium on Parallel Rendering},
  PRS~'95, p. 23–30. Association for Computing Machinery, New York, NY, USA,
  1995.

\bibitem{liukwan1997scalableparallelcell}
K.-L. Ma and T.~Crockett.
\newblock A scalable parallel cell-projection volume rendering algorithm for
  three-dimensional unstructured data.
\newblock In {\em Proceedings IEEE Symposium on Parallel Rendering}, PRS~'97,
  pp. 95--104, 1997.

\bibitem{marmitt2008cpuvolume}
G.~Marmitt, H.~Friedrich, and P.~Slusallek.
\newblock Efficient {CPU}-based volume ray tracing techniques.
\newblock {\em Computer Graphics Forum}, 27(6):1687--1709, 2008.

\bibitem{marmitt2006fastraytet}
G.~Marmitt and P.~Slusallek.
\newblock Fast ray traversal of tetrahedral and hexahedral meshes for direct
  volume rendering.
\newblock In {\em Eurographics /IEEE VGTC Symposium on Visualization},
  EUROVIS~'06, pp. 235--242. The Eurographics Association, 2006.

\bibitem{marsaglia2018explorativevis}
N.~Marsaglia, S.~Li, and H.~Childs.
\newblock Enabling explorative visualization with full temporal resolution via
  in situ calculation of temporal intervals.
\newblock In R.~Yokota, M.~Weiland, J.~Shalf, and S.~Alam, eds., {\em High
  Performance Computing}, pp. 273--293. Springer International Publishing,
  Cham, 2018.

\bibitem{optix}
\mbox{NVIDIA Corporation}.
\newblock {NVIDIA OptiX Ray Tracing Engine}.
\newblock Available at \url{https://developer.nvidia.com/optix}, Accessed: 27
  March 2022.

\bibitem{Fun3DManual}
\mbox{The National Aeronautics and Space Administration (NASA)}.
\newblock {Fun3D Manual}.
\newblock Available at \url{https://fun3d.larc.nasa.gov/}, Accessed: 24 March
  2022.

\bibitem{moran2020fun3d}
P.~J. Moran.
\newblock {FUN3D Retropropulsion Data Portal-NASA}, Jul 2020.
\newblock Available at \url{https://data.nas.nasa.gov/fun3d/}, Accessed: 27
  March 2022.

\bibitem{moreland2011icet}
K.~Moreland.
\newblock {IceT} users’ guide and reference.
\newblock Technical report, Sandia National Laboratories, 2011.

\bibitem{icet}
K.~Moreland, W.~Kendall, T.~Peterka, and J.~Huang.
\newblock An image compositing solution at scale.
\newblock In {\em Proceedings of 2011 International Conference for High
  Performance Computing, Networking, Storage and Analysis}, SC '11. Association
  for Computing Machinery, New York, NY, USA, 2011. doi: {{%
10\hspace{.1pt}\discretionary{.}{%
}{.}\hspace{.4pt}1145\discretionary{/}{%
}{/}2063384\hspace{.1pt}\discretionary{.}{%
}{.}\hspace{.4pt}2063417}}


\bibitem{morrical2019spaceskip}
N.~Morrical, W.~Usher, I.~Wald, and V.~Pascucci.
\newblock Efficient space skipping and adaptive sampling of unstructured
  volumes using hardware accelerated ray tracing.
\newblock In {\em Proceedings of IEEE Visualization}, VIS~'19, pp. 256--260,
  2019.

\bibitem{morrical2020rtxpointlocext}
N.~Morrical, I.~Wald, W.~Usher, and V.~Pascucci.
\newblock Accelerating unstructured mesh point location with {RT} cores.
\newblock {\em IEEE Transactions on Visualization and Computer Graphics}, pp.
  1--14, In Press.

\bibitem{muigg2011interactivevol}
P.~Muigg, M.~Hadwiger, H.~Doleisch, and E.~Groller.
\newblock Interactive volume visualization of general polyhedral grids.
\newblock {\em IEEE Transactions on Visualization and Computer Graphics},
  17(12):2115--2124, 2011.

\bibitem{nelson2006isosurface}
B.~Nelson and R.~M. Kirby.
\newblock Ray-tracing polymorphic multidomain spectral/hp elements for
  isosurface rendering.
\newblock {\em IEEE Transactions on Visualization and Computer Graphics},
  12(1):114–125, Jan. 2006.

\bibitem{pharr2021pbrtaccel}
M.~Pharr, W.~Jakob, and G.~Humphreys.
\newblock Primitives and intersection acceleration.
\newblock In {\em Physically Based Rendering: From Theory To Implementation},
  chap.~4, pp. 169--225. Morgan Kaufmann, Burlington, MA, 2021.

\bibitem{Porter84}
T.~Porter and T.~Duff.
\newblock Compositing digital images.
\newblock {\em ACM Computer Graphics (Proceedings of SIGGRAPH~'84)},
  18(3):253–259, July 1984.

\bibitem{rathke2015simd}
B.~Rathke, I.~Wald, K.~Chiu, and C.~Brownlee.
\newblock {SIMD} parallel ray tracing of homogeneous polyhedral grids.
\newblock In {\em Proceedings of the Eurographics Symposium on Parallel
  Graphics and Visualization}, EGPGV~'15, p. 33–41. The Eurographics
  Association, 2015.

\bibitem{sahistan2021Shell}
A.~Sahistan, S.~Demirci, N.~Morrical, S.~Zellmann, A.~Aman, I.~Wald, and
  U.~G\"ud\"ukbay.
\newblock Ray-traced shell traversal of tetrahedral meshes for direct volume
  visualization.
\newblock In {\em Proceedings of the IEEE Visualization Conference-Short
  Papers}, VIS~'21, pp. 91--95, Oct 2021.

\bibitem{VTK}
W.~Schroeder, K.~Martin, and B.~Lorensen.
\newblock {\em The Visualization Toolkit}.
\newblock Kitware, 4 ed., 2006.

\bibitem{shih2014out}
M.~Shih, Y.~Zhang, K.-L. Ma, J.~Sitaraman, and D.~Mavriplis.
\newblock Out-of-core visualization of time-varying hybrid-grid volume data.
\newblock In {\em 2014 IEEE 4th Symposium on Large Data Analysis and
  Visualization (LDAV)}, pp. 93--100. IEEE, 2014.

\bibitem{shirley1990rastertet}
P.~Shirley and A.~Tuchman.
\newblock A polygonal approximation to direct scalar volume rendering.
\newblock {\em ACM Computer Graphics (Proceedings of SIGGRAPH~'90)},
  24(5):63–70, 1990.

\bibitem{szirmaykalos2011freeps}
L.~Szirmay-Kalos, B.~T{\'o}th, and M.~Magdics.
\newblock Free path sampling in high resolution inhomogeneous participating
  media.
\newblock {\em Computer Graphics Forum}, 30, 2011.

\bibitem{usher2019distributedfb}
W.~Usher, I.~Wald, J.~Amstutz, J.~Günther, C.~Brownlee, and V.~Pascucci.
\newblock Scalable ray tracing using the distributed framebuffer.
\newblock {\em Computer Graphics Forum}, 38(3):455--466, 2019.

\bibitem{Vo2007irun}
H.~T. Vo, S.~P. Callahan, N.~Smith, C.~T. Silva, W.~Martin, D.~Owen, and
  D.~Weinstein.
\newblock {iRun}: {Interactive} {Rendering} of large {Unstructured} grids.
\newblock In J.~M. Favre, L.~P. Santos, and D.~Reiners, eds., {\em Eurographics
  Symposium on Parallel Graphics and Visualization}. The Eurographics
  Association, 2007. doi: {{%
10\hspace{.1pt}\discretionary{.}{%
}{.}\hspace{.4pt}2312\discretionary{/}{%
}{/}EGPGV\discretionary{/}{%
}{/}EGPGV07\discretionary{/}{%
}{/}093\discretionary{%
}{-}{-}100}}


\bibitem{wald2020owl}
I.~Wald, N.~Morrical, and E.~Haines.
\newblock {OWL -- The Optix 7 Wrapper Library}, 2020.
\newblock Available at \url{https://github.com/owl-project/owl}, Accessed: 27
  March 2022.

\bibitem{wald2022bigmesh}
I.~Wald, N.~Morrical, and S.~Zellmann.
\newblock A memory efficient encoding for ray tracing large unstructured data.
\newblock {\em IEEE Transactions on Visualization and Computer Graphics},
  28(1):583--592, 2022.

\bibitem{wald2019rtxpointloc}
I.~Wald, W.~Usher, N.~Morrical, L.~Lediaev, and V.~Pascucci.
\newblock {RTX} beyond ray tracing: Exploring the use of hardware ray tracing
  cores for tet-mesh point location.
\newblock In {\em Proceedings of High-Performance Graphics - Short Papers},
  HPG~'19. The Eurographics Association, July 2019.

\bibitem{Wang2020noveladaptivesampling}
H.~Wang, G.~Xu, X.~Pan, Z.~Liu, R.~Lan, and X.~Luo.
\newblock A novel ray-casting algorithm using dynamic adaptive sampling.
\newblock {\em Wireless Communications and Mobile Computing}, 2020(Article no.
  8822624, 12 pages), 2020.

\bibitem{weiler2003hardwarebased}
M.~Weiler, M.~Kraus, M.~Merz, and T.~Ertl.
\newblock Hardware-based ray casting for tetrahedral meshes.
\newblock In {\em Proceedings of IEEE Visualization}, VIS~'03, pp. 333--340,
  2003.

\bibitem{kuhlen2011libsim}
B.~Whitlock, J.~M. Favre, and J.~S. Meredith.
\newblock Parallel in situ coupling of simulation with a fully featured
  visualization system.
\newblock In T.~Kuhlen, R.~Pajarola, and K.~Zhou, eds., {\em Proceedings of the
  Eurographics Symposium on Parallel Graphics and Visualization}, PGV~'17. The
  Eurographics Association, 2011.

\bibitem{yamaoka2019adaptivetimestepinsitu}
Y.~Yamaoka, K.~Hayashi, N.~Sakamoto, and J.~Nonaka.
\newblock In situ adaptive timestep control and visualization based on the
  spatio-temporal variations of the simulation results.
\newblock In {\em Proceedings of the Workshop on In Situ Infrastructures for
  Enabling Extreme-Scale Analysis and Visualization}, ISAV '19, p. 12–16.
  Association for Computing Machinery, New York, NY, USA, 2019.

\end{thebibliography}
\end{document}